\documentclass[a4paper,fleqn,10pt]{cas-dc}
\usepackage[authoryear]{natbib}
\usepackage{amsmath}
\usepackage{longtable}
\usepackage{graphicx}
\usepackage[inkscapelatex=false]{svg}
\usepackage{hyperref}
\usepackage{colortbl}
\usepackage{subfig}
\usepackage{caption}
\usepackage{float}
\usepackage{cuted}
\usepackage{etoolbox}

\usepackage[T1]{fontenc}
\usepackage[utf8]{inputenc}

\def\tsc#1{\csdef{#1}{\textsc{\lowercase{#1}}\xspace}}
\tsc{WGM}
\tsc{QE}
\def\userfigwidth{\columnwidth}

\begin{document}
\let\WriteBookmarks\relax
\def\floatpagepagefraction{1}
\def\textpagefraction{.001}

\shorttitle{67P/C-G gravity from Rosetta's Doppler and optical data}    

\shortauthors{JLV, TJ, JCM, LJ, SLM, RG}  

\title [mode = title]{New gravity field of comet 67P/C-G based on Rosetta's Doppler and optical data}  



\author[1]{Julien Laurent-Varin}

\cormark[1]

\ead{julien.laurent-varin@cnes.fr}

\credit{Develop and maintain the GINS software, introduces the "landmark" measurement function into GINS, carried out the simulations and produced the results of the study}

\affiliation[1]{organization={CNES},
            addressline={18, avenue Edouard Belin}, 
            city={Toulouse},
            postcode={F-31401}, 
            country={France}}
            
\cortext[1]{Corresponding author}

\author[1, 2]{Théo James}



\credit{Carried out the simulations and produced the results of the study}

\affiliation[2]{organization={ESA},
            addressline={Keplerlaan 1}, 
            city={Noordwijk},
            postcode={2201 AZ}, 
            country={Netherlands}}

\author[1]{Jean-Charles Marty}




\author[3]{Laurent Jorda}



\affiliation[3]{organization={Aix Marseille Univ},
            addressline={CNRS, CNES, LAM}, 
            city={Marseille},
            country={France}}

\author[4,5]{Sebastien Le~Maistre}


\affiliation[4]{organization={Royal Observatory of Belgium},
            addressline={Avenue Circulaire 3}, 
            city={ Uccle},
            postcode={BE-1180}, 
            country={Belgium}}
\affiliation[5]{organization={Université Catholique de Louvain},
            addressline={Place Louis Pasteur 3}, 
            city={Louvain-La-Neuve},
            postcode={BE-1348}, 
            country={Belgium}}

\author[6]{Robert Gaskell}


\affiliation[6]{organization={Planetary Science Institute},
            addressline={1700 East Fort Lowell, Suite 106}, 
            city={Tucson},
            postcode={AZ 85719-2395}, 
            country={USA}}


\begin{abstract}
The gravity field of a celestial body gives valuable insights into its fundamental properties such as its density and internal structure. The Doppler data collected by the Radio-Science Investigation (RSI) experiment of the Rosetta mission were previously used to determine the gravity field of comet 67P/Churyumov–Gerasimenko up to degree 2 \citep{Patzold_2016}. In the present study we re-estimate the gravity field of 67P/C-G using not only RSI data as before, but also images data from Rosetta's OSIRIS camera. These data, converted into "landmark" observations, are complementary to RSI data. 
Therefore, the analysis of combined Doppler and optical data results in a significant improvement in the restitution of Rosetta's orbit and the determination of the comet gravity field with respect to previous work. Some coefficients of the comet's gravity field are now resolved up to degree 4. The mass and low degrees estimates are in fairly good agreement with those previously published, but the improvement in their accuracy (i.e. lower sigmas) as well as the better resolution (i.e. maximum degree) of the new gravity field suggests that the distribution of mass in the nucleus may not be uniform, contrary to what was previously thought. 
Moreover, we estimate a change in the mass of the comet attributed to ice sublimation at its orbital perihelion that is almost three times greater than that previously published. The new estimated mass loss is $\Delta M=28.0 \pm 0.29 \times 10^9 kg$, corresponding to $0.28\%$ of the total mass of the comet. Thanks to a precise determination of the degree-1 gravity coefficients, we observe for the first time a motion of the center of mass of the comet by $\sim35\,m$ northward that could be explained by a more pronounced outgassing activity in the south of the comet due to the orientation of its spin axis relative to the Sun. 
The temporal evolution (before versus after perihelion) of the other estimated gravity coefficients and in particular degree-2 is more modest ($0.8\%$ for $C_{20}$ and $2\%$ for $C_{22}$, $S_{22}$). 
\end{abstract}



\maketitle


\section{Introduction}\label{sec:intro}

After a 10-year cruise, the Rosetta spacecraft \citep{Kolbe1997,Glassmeier2007} arrived at comet 67P/C-G (i.e., 67P/Churyumov–Gerasimenko) on August 6th, 2014, and orbited around its oddly-shaped nucleus for a little more than two years, until September 30th, 2016. The measurements acquired during that time are invaluable for exploring the fundamental characteristics of the nucleus (mass, density, porosity, composition). Among others, the Radio Science Investigation Experiment (RSI, \cite{Patzold_2007}) onboard Rosetta was dedicated to these topics (These data are available at The European Space Agency's Planetary Science Archive (PSA)\footnote{https://www.cosmos.esa.int/web/psa/rosetta}) 

The exploitation of Doppler measurements led to many important results (see \cite{Patzold_2016,Patzold_2019}). First of all, the mass of the nucleus was accurately determined ($GM = 666.2 \pm 0.2 m^3.s^{-2}$). Combined with an accurate shape model of the comet, this allowed us to compute the average density of the body, providing insights into the porosity and overall dust-to-ice mass ratio \cite{Patzold_2016}. Secondly, the mass loss between 2014 and 2016, resulting from the sublimation of ice during the perihelion passage (in Aug. 2015), was first estimated by \cite{Patzold_2019} to $\Delta M=10.5 \pm 3.4 \times 10^9 kg$, corresponding to $0.1\%$ of the total mass. This measurement was important to constrain the dust-to-gas and refractory-to-ice mass ratios, as discussed in \cite{Choukroun_2020}, and revealed that most of the evaporated mass eventually fell back on 67P/C-G, leading to a global mass redistribution throughout the surface of the comet.
Finally, the degree-2 spherical harmonic coefficients of the gravity field have been estimated by \cite{Patzold_2016} with statistical significance for $\bar{C}_{20}$ and $\bar{C}_{22}$. These values can be compared to those derived from shape models in order to deduce the level of heterogeneity inside the nucleus.

Several high-resolution shape models have been reconstructed from NAVCams and/or OSIRIS images \citep{Jorda_2016,Preusker_2017,ESA_shape_hires}. We determined that all of them perfectly agree for the first few degrees of the gravity field, the difference remaining well below the formal errors of the spherical harmonics gravity coefficients.
We therefore choose to use the shape model of \cite{Jorda_2016}, from which one can compute the gravity field of a homogeneous comet, where the mass is uniformly distributed across the nucleus. The position of the center of mass calculated with this hypothesis is shifted by $(18 \pm 7, -32 \pm 4\ \mathrm{m}, 16 \pm 10\ \mathrm{m})$ with respect to the actual center of mass of the comet \citep{Jorda_2016}, suggesting an inhomogeneous density distribution \citep{Davidsson_2016}.
The analysis of \cite{Patzold_2016}, based on Rosetta Doppler measurements only, suggests that the mass is uniformly distributed (coefficients of a homogeneous comet are within the confidence interval of their estimated counterparts). However, several other studies, based on different kinds of data, suggest a non-uniform density distribution. For instance, the analysis of CONSERT data shows that the sub-surface (up to about 25~m) around the final landing site of the Philae lander is significantly denser than the deeper part of the nucleus \citep{Kofman_2020}. Also, the analysis of the excited rotational state detected during the shape reconstruction \citep{Preusker_2015,Jorda_2016} indicates that a uniform density is not compatible with the measured rotation and precession periods of the comet \citep{Gutierrez_2016}.
Finally, a detailed three-dimensional model of the layers identified in the two lobes suggests that the small lobe was compressed during the impact which led to the formation of the bilobate nucleus of 67P \citep{Franceschi_2019}. This would imply that the more compact neck region would be denser than the two lobes.

Most of these small inhomogeneities have a signature in the gravity field which is too small to be observed in the measured field given its accuracy and resolution. Reducing uncertainties and increasing the degree of the spherical harmonic expansion of the field is the only way to detect mass anomalies from the orbit.
In fact, to some extent, such an improvement should still be possible using Rosetta data. Indeed, because, on occasion, the probe was as close as 7~km from the center of the nucleus, it is reasonable to assume that Rosetta's orbit is sensitive to degrees of the gravity field higher than 2. Furthermore, Doppler measurements are not the only kind of data that can be used for the Precise Orbit Determination (POD). Indeed, one can also use images of the comet to constrain the spacecraft position at the time of their acquisition, since such images can be reverted into positions of Rosetta in the nucleus frame. 

Several tens of thousands of images of the nucleus acquired by the OSIRIS camera \citep{Keller_2007} have been used by the SPC software developed by \cite{Gaskell_2008} to reconstruct the global shape of the nucleus of 67P \citep{Jorda_2016}. During the process, a huge set of stereo landmarks are defined at the surface of the comet, and their coordinates in the images as well as in the body-fixed frame are calculated.
These measurements are complementary to Doppler data. While Doppler measurements constrain the velocity along the line-of-sight (Earth-spacecraft direction), OSIRIS images constrain the position of the spacecraft relative to the comet. In other words, the Doppler measurements contain information about the speed of the spacecraft (in the line-of-sight) while the optical observations are anchors for the relative position of the comet on the trajectory. Landmarks have been increasingly used in the field of planetary geodesy for the last two decades. For small bodies especially, it has been found to efficiently decrease uncertainties that are typically rather high due to the low gravity: for instance for asteroids Eros (\citep{Konopliv_2002}, mission NEAR), Vesta (\citep{Konopliv_2014}, mission DAWN), or Bennu (\citep{Goossens2021}, mission OSIRIS-REx). In the case of the Rosetta mission, optical data have been used for navigation purposes \citep{Godard2016MultiarcOD} and also in the scientific study published by \cite{GAO2023}. The latter estimated the gravity field coefficients up to degree 3 based on pre-perihelion data only. The data they used consist in both radiometric observations and NAVCAM images (less accurate than the OSIRIS scientific camera). \cite{GAO2023} did not model the comet's outgassing, nor the degree 1 impact, and they (obviously) didn't estimate different coefficients before and after perihelion.

Based on the above, we re-estimate the gravity field of comet 67P/C-G up to the highest degree achievable, using both Doppler and optical measurements. The GINS (see \cite{doc_algo_GINS}) and DYNAMO software of the French Centre National d'Etude Spatiales (CNES) are used for these calculations. In section \ref{sec:data} we present our modeling work and the observations we use to fit the model. In section \ref{sec:method} we detail the estimation methodology, while in section \ref{sec:results} we show our results and in section \ref{sec:discussion} we discuss the implications for the mass distribution within the nucleus as well as the composition of the coma.

\section{Data and models} \label{sec:data}

This section summarizes our model of the spacecraft's dynamics, and present the  observations used in this study.

\subsection{Periods of interest}

\begin{figure*}[ht]
     \centering
     \includegraphics[width=\textwidth]{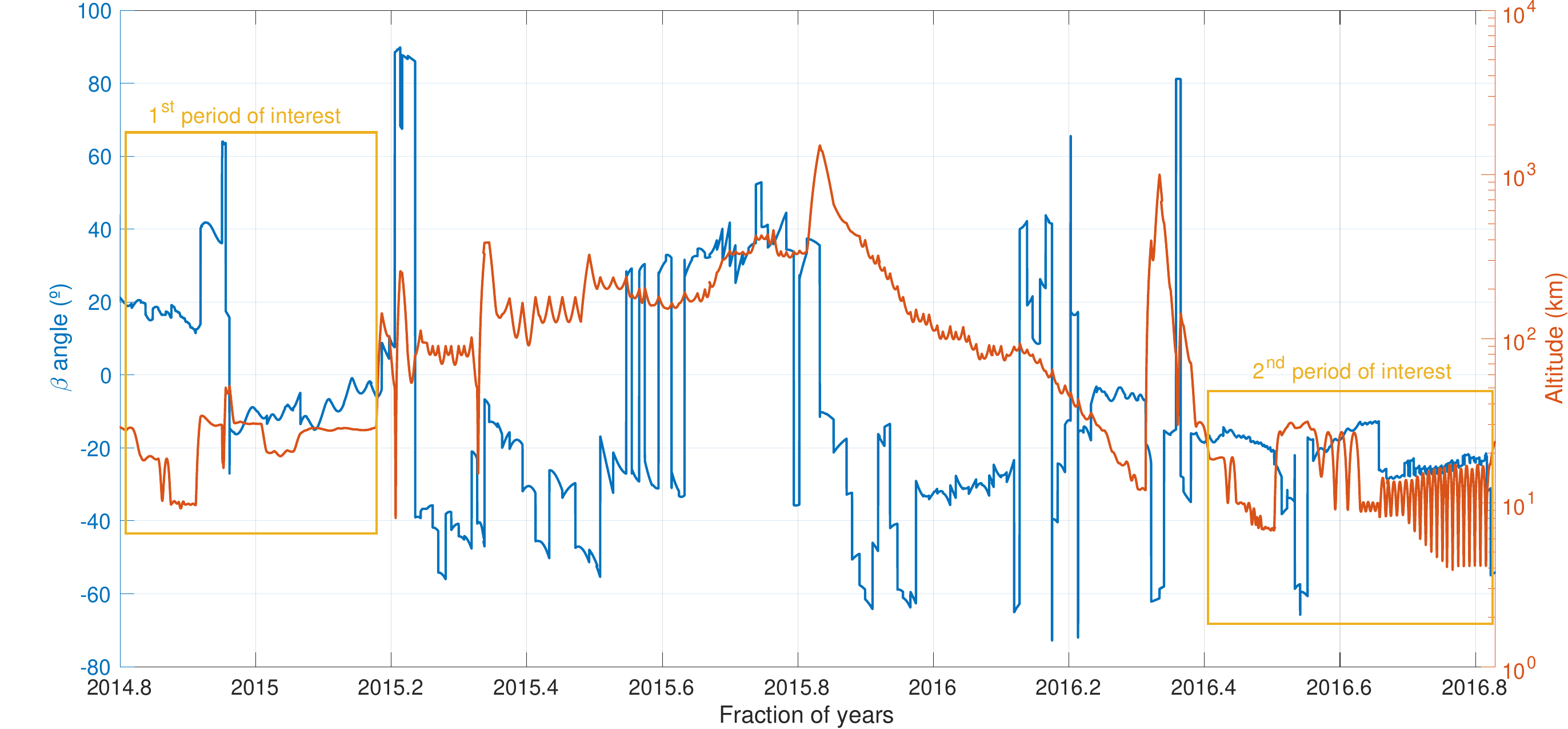}
     \caption{Evolution of the spacecraft-comet distance (right y-axis) and the angle between the orbit's normal vector and Earth direction, so called $\beta$ angle (left y-axis), over the course of the mission}
     \label{fig:beta}
\end{figure*}

We focus our analysis on two periods of the mission: one in late 2014/early 2015 before perihelion and one in mid-2016 after perihelion. During these two periods the sensitivity to 67P gravity is maximum since the distance from Rosetta to the center of the comet is minimum, most of the time smaller than 30~km, and sometimes as low as 7~km (see Fig.~\ref{fig:beta}). In addition, the out-gassing activity of the comet is inversely proportional to the distance to the sun, therefore it was maximum between these periods. From a gravity estimation perspective, high out-gassing means that significant aerodynamic forces perturb the spacecraft trajectory and make the POD procedure too complex and uncertain. That's also why, for mission safety, the probe was moved away from the comet during the passage at perihelion, de facto reducing the sensitivity of the orbit to the gravity field. The combination of high out-gassing and large distance to the comet led us to exclude the perihelion period from February 2015 to April 2016 from our analysis. This leaves us with 132~days of data with a good sensitivity to the comet's gravity field. 
Before any editing and studies, we start with 36 arcs for the pre-perihelion period and 60 arcs for the post-perihelion period. At the end of study we kept 15 arcs pre-perihelion and 43 post-perihelion. Details of the arcs kept are given in the appendix (see Appendix~\ref{sec:arcs}).

\subsection{Modeling} \label{sec:modeling}

This sub-section brings together the physical modelling details and presents the forces and accelerations taken into account in the trajectory calculation. They include the gravity field of the central body (67P/C-G), Center of Mass offset, out-gassing accelerations, Solar-Radiation-Pressure (SRP) and third body accelerations (Sun, Jupiter, Earth, Moon, Venus, ...).

\subsubsection{Shape}
\label{sec:shape}

For the global shape of the nucleus, we use the latest SPC model \citep{Jorda_2016}
\footnote{The model labeled ``SPC shap8 v2.1''.} at a resolution of about 4~m (i.e., with 3.1 millions of facets).
The model is based on about 49,000 OSIRIS images (both NAC and WAC) which are analyzed to retrieve the topography of the nucleus.
The reconstruction is split into a total of about 25,000 small topographic units called ``maplets''
\footnote{SPC maplets are squared elevation models of 99x99 elements, at a typical resolution of 1~m.} covering the entire surface of the comet.
During the analysis, a stereo landmarks is defined at the center of each maplet.
These landmarks are used in SPC to retrieve the geometric information associated to each image.
The SPC analysis provides the body frame and image coordinates of the landmarks as well as the camera position and pointing direction for each image analyzed by the software.
All SPC data (shape and landmarks coordinates) are calculated in the ``Cheops'' reference frame \citep{Preusker_2015}.\\
Other more accurate shapes exist  (e.g. \cite{Chen2023}), but we chose not to use them firstly to use a shape model consistent with the landmarks definition, and secondly because the shape is used in our method only to initialise the gravity field under the assumption of uniform mass distribution. The resolution of the shape is significantly higher than that of the gravity field (even at degree 20), so the gravity field calculated from a more precise shape model gives essentially identical Stokes coefficients.

For the sake of clarity, we define three specific center points here: the Centre of Mass (CoM), the Centre of Reference (CoR) and the Centre of Figure (CoF). The CoM is the body's true physical centre of gravity, i.e. the barycentre of all the masses. The CoR is the centre of the frame of reference in which the landmarks and spherical harmonics of gravity are described, i.e. the Cheops frame of reference. Finally, CoF is the centre of gravity if the distribution of masses were uniform in the shape.

\subsubsection{67P/C-G ephemeris, rotation and orientation}
\label{ssec:ephrotor}

The ephemeris, orientation and rotation of the comet are extracted from the latest SPICE kernels \citep{SPICE_2019} reported in Tab.~\ref{tab:kernels}.
As stated in the SPICE documentation, the orientation of 67P/C-G cannot be represented over a long period of time using the standard IAU formulation. Instead, it is recommended to use attitude kernels (CK)  provided by the mission and archived in the SPICE repository to orient the comet over time. Thus, a discretisation of the comet's orientation has been constructed in the form of a quaternion table, and supplied to GINS.

\begin{table*}[ht]
\caption{List of SPICE kernels used in the simulation}\label{tab:kernels}
\centering
\begin{tabular}{c|l|l}
\toprule
Type & Description & Kernel name \\
\midrule
CK & 67P/C-G attitude kernel (pre-perihelion)  & \textsf{\detokenize{CATT_DV_145_02_______00216.BC}} \\
CK & 67P/C-G attitude kernel (post-perihelion) & \textsf{\detokenize{CATT_DV_257_03_______00344.BC}} \\
FK & Rosetta spacecraft frame kernel           & \textsf{\detokenize{ROS_V38.TF}} \\
LSK & Leap seconds kernel                 & \textsf{NAIF0012.TLS} \\
SPK & Comet 67P/C-G ephemeris             & \textsf{\detokenize{CORB_DV_257_03___T19_00345.BSP}} \\
SPK & Planetary ephemeris                 & \textsf{DE440.BSP} \\
SPK & Rosetta spacecraft trajectory       & \textsf{\detokenize{RORB_DV_257_03___T19_00345.BSP}} \\
\bottomrule
\end{tabular}
\end{table*}

\begin{figure*}[ht]
	\centering
    \includegraphics[width=\userfigwidth]{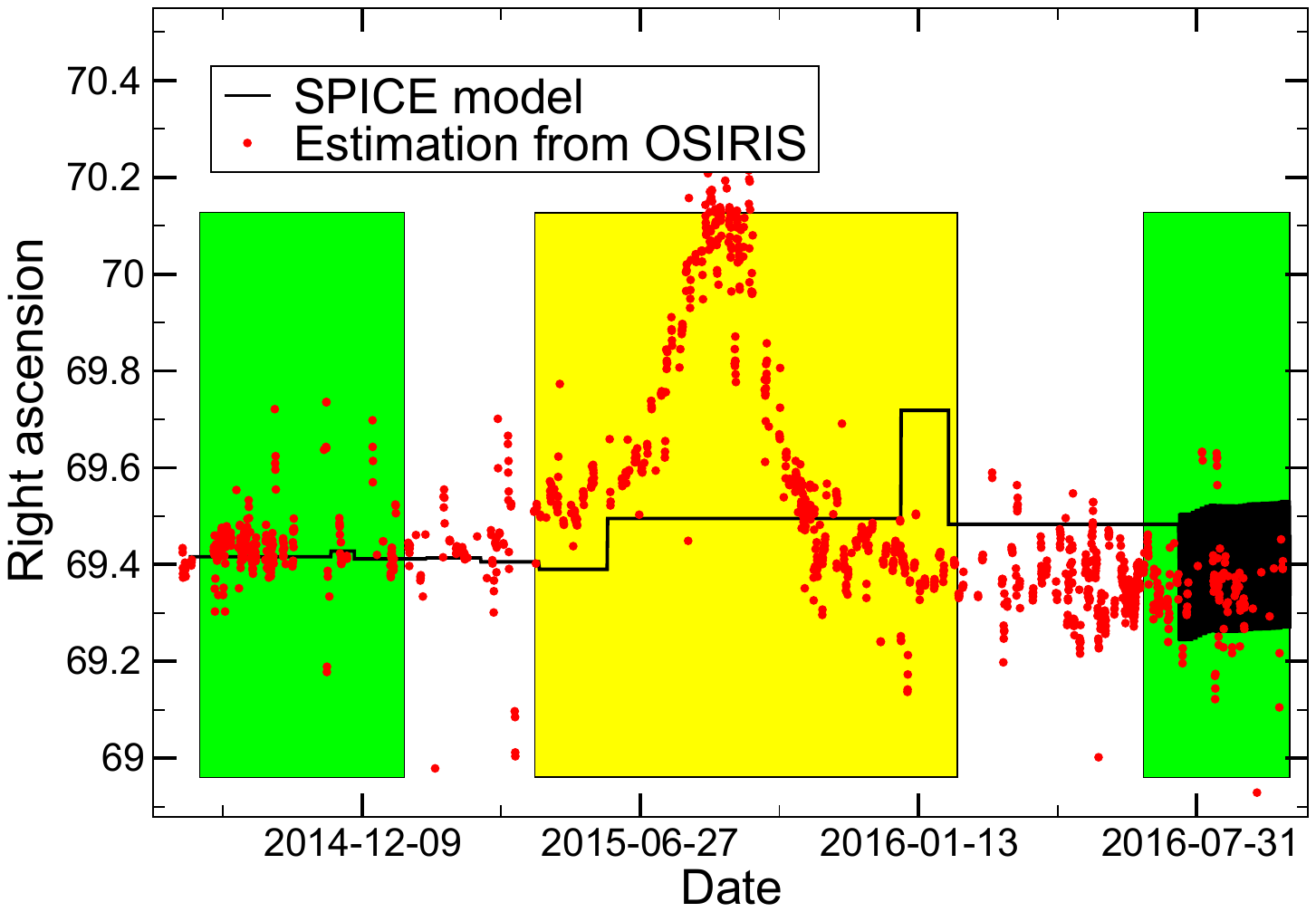}
    \includegraphics[width=\userfigwidth]{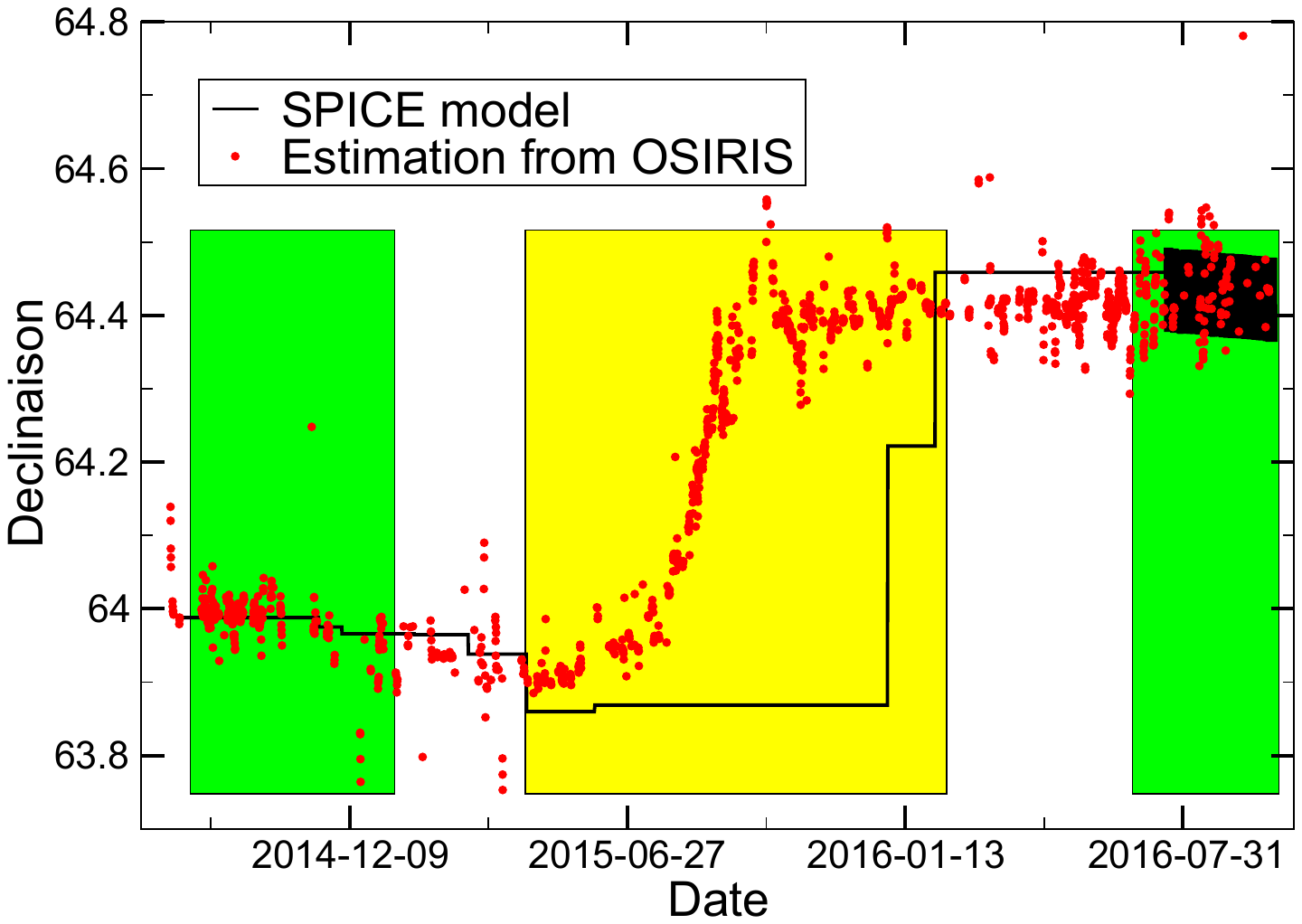}
    \caption{ Right ascension (left) and declination (right) of CG/67P comparison from SPICE and from OSIRIS restitution. The green zones are the periods of interest, and the yellow zone is the perihelion period.}
    \label{fig:AD_CG67P}
\end{figure*}

The model of orientation of the spin axis of 67P in right ascension and declination (Fig.~\ref{fig:AD_CG67P}) based on SPICE kernels (Tab.~\ref{tab:kernels}) are compared to those resulting  from OSIRIS picture analysis \citep{Jorda_2016}. We observe piece-wise constant values in the SPICE model for the first part of the mission, while a more precise model is given after July 2016. The differences between the SPICE model and the OSIRIS reconstructions suggest that the comet's orientation is accurate to between a tenth of a degree and half a degree. Comet orientation errors can induce errors in the estimate of the gravity coefficients which we avoid here by adding landmark data (helping to position the spacecraft in the body-fixed frame and therefore relative to its gravity potential) and adjusting the pointing of the OSIRIS camera.

As for the ephemeris of the comet, we used the one recommended by ESA because it showed better performance (i.e. smaller post-fit residuals, more arcs that converge) than the one provided by JPL \citep{Farnocchia2021}. We think that the lower performance obtained with the JPL ephemeris is inherent to the way it is constructed, fitting not only pseudo-distance measurements derived at given epochs of the Rosetta mission, but also terrestrial astrometric measurements. The so-obtained continuous orbit over a long period is therefore the best compromise between all these measurements but does not specifically optimise the trajectory of the comet at the time of Rosetta's 2-way Doppler measurements acquisition, as does the ESA ephemeris. It should also be pointed out that \cite{Farnocchia2021} has only used the NASA DSN measurements in its processing, and not the ESTRACK ESA measurements, which are more numerous, and the optical measurements from NAVCAM, which are less accurate than the OSIRIS instrument.

\subsubsection{Manoeuvres and wheel off-loadings}
The trajectory of the spacecraft was controlled with regular manoeuvres, generating substantial velocity increments ($\Delta V$). The magnitude of those $\Delta V$ is much larger than the gravity force and their values are not accurately known. Estimating them would be a hazardous task, therefore we design our arcs to exclude these maneuvers, which occur every few days. The exclusion of these maneuvers have been the driver of our arc splitting procedure, leading to arcs duration between 16 hours and 150 hours (see Tabs.~\ref{tab:arcs_def_14} and~\ref{tab:arcs_def_16}).\\

\noindent
The spacecraft controls its attitude with reaction wheels which need to be periodically desaturated, leading to engine activation called wheel off-loading (WoL) maneuvers. Given the relatively old technologies onboard Rosetta, WoL maneuvers generate significant residual $\Delta V$s, which have to be taken into account in the POD. Since these WoL occur twice a day, it is not possible to avoid them in our computation arcs (the resulting arcs would then be too short to be sensitive enough to the comet's gravity field). Therefore, we estimate their values in the orbit determination process.

\subsubsection{Comet gravity field}
\begin{figure}[ht]
     \centering
     \includegraphics[width=\userfigwidth]{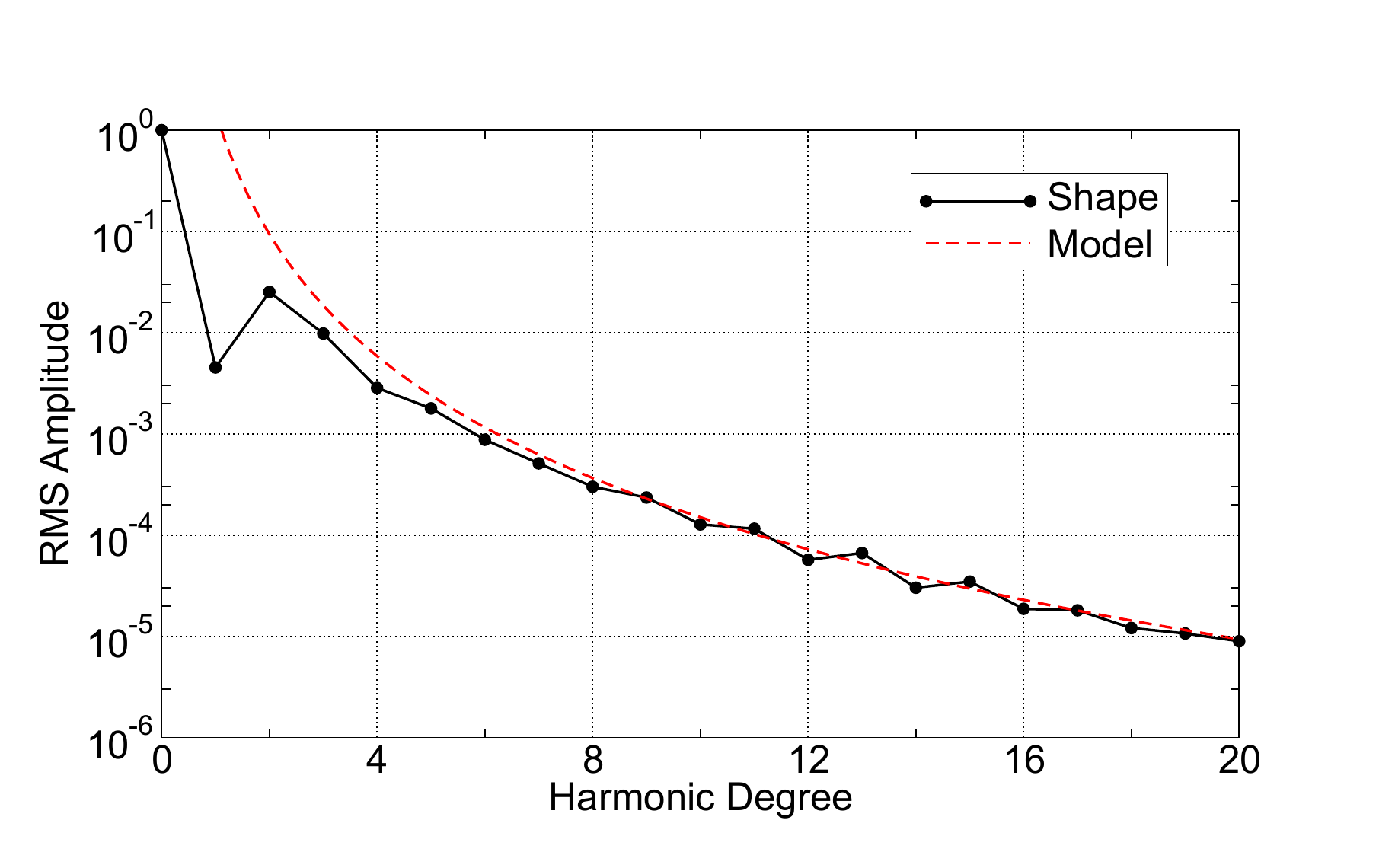}
     \caption{Spectral representation of Stokes coefficients from shape, red dashed line corresponds to $1.5/l^4$.}
     \label{fig:sp_shape}
\end{figure}

Despite its non-spherical shape, we model the gravitational potential of the central body using classical spherical harmonic expansion according to:

\begin{align*}
 U(r,\phi,\lambda) = & \frac{GM}{r} \left[ \bar{C}_{00} +
 \sum^\infty_{l=1} \left(\frac{R}{r}\right)^l \sum^l_{m=0} \bar{P}_{l,m}(\sin\phi) \right. \\
   & \left. \times \Big( \bar{C}_{l,m}\cos {m\lambda} +  \bar{S}_{l,m}\sin{m\lambda} \Big) \right],
\end{align*}


where $r$ is the distance to the comet' CoR (in which spherical harmonic expansion is defined) and $\phi,\lambda$ the latitude and longitude of the field-point, respectively. $G$ is the gravitational constant, $M$ is the total mass of the 67P/C-G and $R$ is the equatorial radius of $2650.0\ \mathrm{m}$, consistent with that of \cite{Patzold_2016}. $\bar{P}_{l,m}$ is the fully normalized associated Legendre function of degree $l$ and order $m$ and $\bar{C}_{l,m},\bar{S}_{l,m}$ are the normalized Stokes coefficients. In this paper, all the reported values for the Stokes coefficients will be normalized. The normalisation factor is classically \citep{Kaula1966} computed as follows

$$N_{lm}=\sqrt{\frac{(2-\delta_{0m})(2l+1)(l-m)!}{(l+m)!}},$$
where the Kronecker delta $\delta_{0,m}$ is equal to 1 for $m=0$, and 0 otherwise.\\
\noindent
 In practice, the spherical harmonic expansion is done up to a given degree ($l_{max}$) which we arbitrarily set to 20. The Fig.~\ref{fig:sp_shape} shows the Root Mean Square (RMS) power spectrum of the gravity field of 67P computed according to \citep{Kaula1966}:.
\begin{equation}
P_l=\sqrt{\frac{\sum_{m=0}^l (\bar{C}_{lm}^2+\bar{S}_{lm}^2)}{2l+1}}, \label{rms:CS} 
\end{equation}
where the $\bar{C}_{lm},\bar{S}_{lm}$ are deduced from the shape of the comet assuming uniform internal mass distribution (see supp.mat.). 
As can be deduced from Fig.~\ref{fig:sp_shape}, the decay of the gravity spectrum follows a power law in $K/l^4$. This is very different from that of most of the other bodies of the solar system that follow Kaula's law in $K/l^2$, simply because 67P/C-G has a very irregular shape that is very different from the spherical shape assumed in Kaula's equation.

\subsubsection{Solar Radiation Pressure}
The acceleration experienced by the spacecraft due to solar radiation pressure depends on the thermo-optical properties of the surfaces that compose it. It is modeled as follows:
\begin{equation}
    a_{SRP} = F_S \frac{\Phi}{c \cdot {d_{SE}}^2} \left( \frac{d_{SE}}{d_{S}} \right)^2 \sum_k S_k \bar{R}_k,\label{eq:SRP}
\end{equation}
where $F_S$ is scale factor of the solar pressure force, $c$ speed of light, $\Phi$ the solar flux at 1 AU (Astronomical Unit : mean distance between sun and earth), $d_{SE}$ one AU, $d_S$ the distance between spacecraft and sun, $S_k$ area of face k of the satellite and $\bar{R}_k$ the reflectivity vectorised coefficient on face k of the satellite (depends on the reflectivity coefficients of the face and one the angle of incidence lighting (without units).  The set of facets defined by their specific shapes, surfaces and thermo-optical coefficients used in this formula compose the so-called macro-model of the spacecraft (A $2.8 m \times 2.1 m \times 2.0 m$ box and two $32.31 m^2$ wings). The SRP is small, but cannot be neglected because the total area of the macro-model of Rosetta is quite big, $64\ \mathrm{m}^2$, which induces a SRP force of the order of a few $10^{-9}\ \mathrm{m}/\mathrm{s}^2$ (see Fig \ref{fig:magn}). 
To account for uncertainties in the spacecraft macro-model, a single SRP scaling factor is estimated for the entire mission. It can vary slightly around one, thus allowing to partially correct for small imperfections of the model.

\subsubsection{Outgassing-induced aerodynamic forces}
Even though the selected arcs avoid periods of intense activity, some residual out-gassing still occurs during our periods of interest and has to be accounted for in the orbit computation. The interaction of the gas with the spacecraft generates aerodynamic forces in the direction of their relative motion. Since the velocity of the spacecraft in the body-fixed frame (few tens of cm/s) is very low compared to the out-gassing velocities (few hundred m/s), the resulting aerodynamic force is mostly centrifugal. Neglecting it would therefore result in a direct error in the gravity force, which is centripetal. We model the out-gassing velocity as \citep{Hansen2016}:

$$ v=\left(-55.5 \cdot r_h + 771\right)\left(1+0.171 \cdot e^{-\frac{r_h-1.24}{0.13}}\right), $$
where $v$ is the gas velocity (assumed to be radial only, oriented outwards) and $r_h$ is the heliocentric distance of the nucleus. The numerical values are empirical constants resulting from the estimation of the gas density carried out using the ROSINA instrument. Indeed, the COPS experiment of ROSINA yields an estimate of the density of molecules around the nucleus (\citep{ROSINA_prl,ROSINA_esc1,ROSINA_esc2,ROSINA_esc3,ROSINA_esc4,ROSINA_ext1,ROSINA_ext2,ROSINA_ext3}). 
Assuming that the mean molecular mass of these particles is the mass of water vapour $H_2O$ (18 g/mol), we can then build a gas model as a radial wind emanating from the comet with a measured mass density. While this force can be compared to a drag force, its orientation is fundamentally different.

In \cite{Hansen2016}, it is stated that the velocity model has an error of less than 0.1\%, and that the ROSINA water abundances have an uncertainty of 10\%, meaning that the major part of the error on this aerodynamic force will be of the order of 10\%. It should be emphasised that the contribution of this force is very small.

Nevertheless, the gas flow interacts with each face of the spacecraft described by its macro-model in accordance with the description in the document \cite{doc_algo_GINS}. This acceleration due to aerodynamic forces has an expression similar to the solar radiation pressure acceleration, involving a contribution from each surface of the spacecraft model.

Fig.~\ref{fig:magn} shows the respective magnitude of each force applied on the spacecraft. The outgassing-induced aerodynamic forces are indeed low compared to the gravity or the SRP, but not negligible, especially at low altitude.

\subsubsection{Non inertial reference frame}
\label{sec:deg1}

The dynamic system representing the centre of the spacecraft is integrated in a reference frame assumed to be Galilean (or inertial). By construction, this frame of reference is centered on the CoR of the comet, which may differ from its CoM. The non-zero distance between CoR and CoM undermines the assumption that the integration frame of reference is inertial. An additional acceleration representing the movement of the CoR around the comet's CoM must therefore be taken into account. 
Thus, as soon as the coefficients ($\bar{C}_{11}$,$\bar{S}_{11}$) are different from zero, the integration frame of reference loses its inertial character. In such a case, an additional acceleration $\vec{\gamma}_{1}$  of the following form must be taken into:

$$
 \vec{\gamma}_{1} = -\vec\Omega \times (\vec\Omega \times \overrightarrow{CO}) - \frac{d \vec\Omega}{dt} \times \overrightarrow{CO},
$$
where $C$ is the comet's CoM and O is the CoR. Assuming that the comet's spin angular vector $\vec\Omega$ remains constant during the time periods considered and oriented in the $z$ direction, then the frame acceleration can be expressed as a function of the degree-1 coefficients according to:

$$
\vec{\gamma}_{1} = -\Omega^2 \sqrt{3} R \begin{pmatrix}
 \bar{C}_{11} \\ \bar{S}_{11} \\ 0 
\end{pmatrix}. 
$$
$R=2650.0\ \mathrm{m}$ is the equatorial radius of the nucleus and $\Omega$ is its mean rotation rate equal to $0.14072\ 10^{-3}\ \mathrm{rad}/\mathrm{s}$ for the pre-perihelion period and to $0.14151\ 10^{-3}\ \mathrm{rad}/\mathrm{s}$ for the post-perihelion period (values  from \cite{Godard2016MultiarcOD}).

\noindent
Traditionally set to zero,  we estimate here the degree-1 coefficients, and therefore their impact on the dynamics must be taken into account. For 67P/C-G, we expect these coefficients to have non-zero values varying with time  because of ice sublimation processes which should induce movements of the CoM in the body-fixed frame during perihelion. Moreover, because the landmark data are sensitive to the position of the CoR, while the Doppler data provide information about the position of the CoM, the combination of these two data sets gives us sensitivity to degree-1 that can be used to determine the CoR-CoM offset.\\
Note that the acceleration $\vec{\gamma_1}$ is actually quite significant, of the order of $10^{-9} m.s^{-2}$ given the order of magnitude of our estimates of $\bar{C}_{11}$ and $\bar{S}_{11}$  (see Sec.~\ref{sec:results}), which is comparable to the comet higher degree gravity acceleration (see Fig.~\ref{fig:magn}).

\subsubsection{Accelerations magnitude}
\begin{figure}[ht]
     \centering
     \includegraphics[width=\userfigwidth]{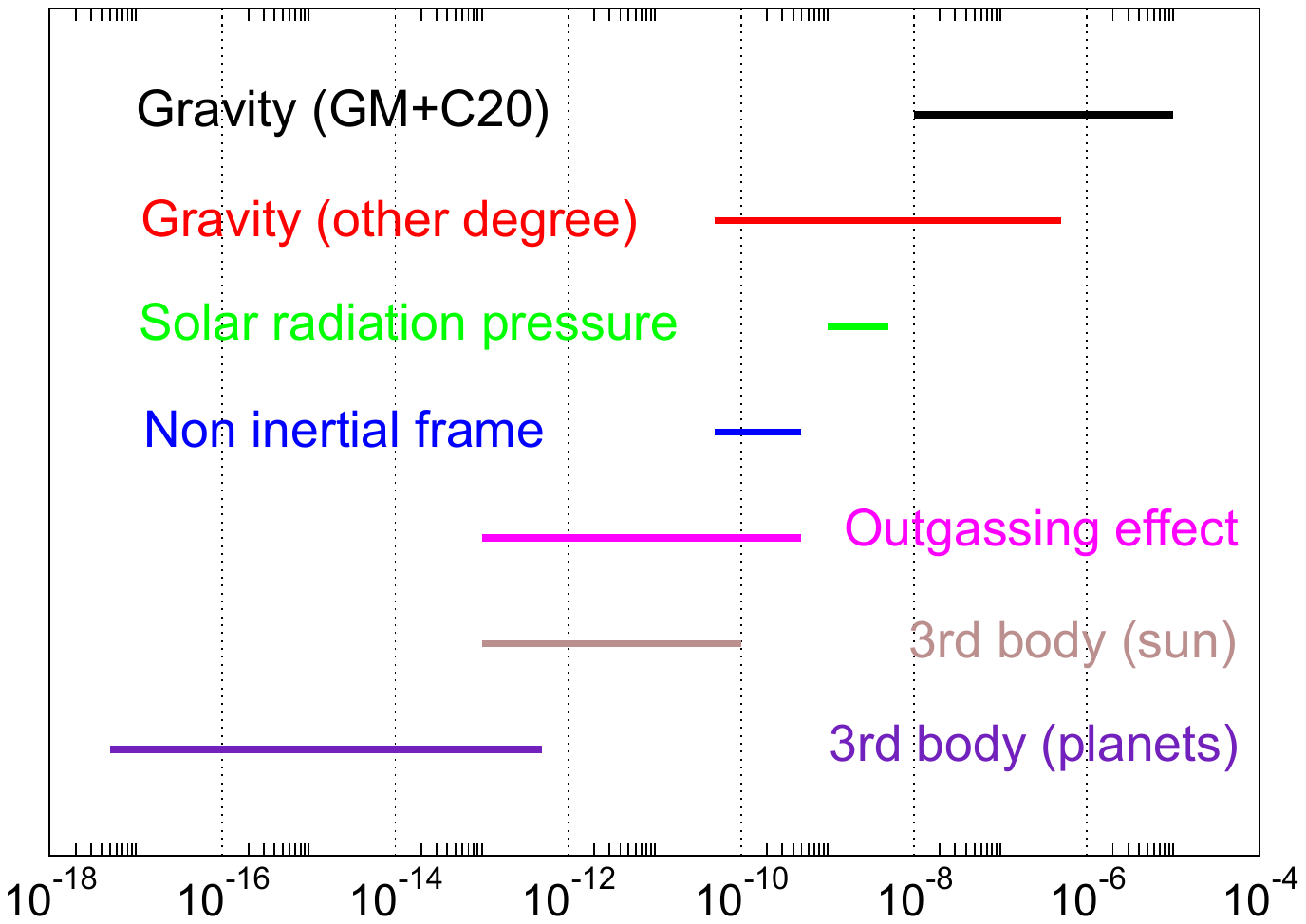}
     \caption{Magnitudes of the accelerations acting on Rosetta $[\mathrm{m}/\mathrm{s}^2]$ during the mission (Distance to coma lower than 30~km)}
     \label{fig:magn}
\end{figure}

The complete dynamical model and the magnitudes of each type of accelerations is presented in Fig.~\ref{fig:magn}. The ranges plotted in this figure encompass the accelerations amplitudes (or norm) of each of the arcs computed in this study.
The gravity field is broken down into two parts: the central contribution plus $\bar{C}_{20}$, and the contribution of the other coefficients. As expected, the former are dominant. Nevertheless, there are also other important forces such as the solar radiation pressure, which thus needs to be precisely modeled. 
As far as gravity is concerned, it should be reminded that the terms of degree $d$ contribute to the acceleration of gravity proportionally to $1/r^{2+d}$. Therefore, reducing the distance to the comet by a factor of 2, increases the central term (of degree 0) acceleration by a factor of 4, that of the terms of degree 1 by a factor of 8 and more generally the accelerations induced by the terms of degree $d$ by a factor of $2^{2+d}$. This explains the large variability of the 67P gravitational accelerations undergone by Rosetta and shown in Fig.~\ref{fig:magn}, as a result from the spacecraft altitude variation (Fig.~\ref{fig:beta}). 
The magnitude of the outgassing-induced aerodynamic forces and the non-inertial frame acceleration are small but comparable to high-degree gravity effects. They can be detected and thus estimated using Rosetta's RSI data. Indeed, an acceleration of $10^{-10} m/s^2$ acting over 6 days (which is the maximum duration of our arcs) can cause the velocity to vary by around $0.05 mm/s$, which corresponds to a signature in the Doppler signal of $2.5 mHz$, which is the measurement noise. Therefore, all the forces inducing acceleration above this threshold of $10^{-10} m/s^2$ are estimated while those below that threshold are kept fixed.

\subsection{Observations}
\subsubsection{Doppler measurements}
Doppler tracking measurements are collected for the Rosetta mission as part of the RSI experiment. The data used in this study are two-way X-band Doppler observations averaged over 60 seconds (See \cite{Moyer_2005}) and collected by the ESTRACK antennas located in New Norcia (Australia), Cerebros (Spain) and Malargüe (Argentina). After data calibration and editing (see Sec.~\ref{subsec:data_qual}), the nominal level of noise is typically $\sigma_{DOP} = 3\ \mathrm{mHz}$.  \\
Such a kind-of-limited Doppler precision is partly due to the orientation of Rosetta's orbital plane, which remains pretty low over the periods of interest, especially pre-perihelion where the orbit is close to a face-on configuration (beta angle around $20^{\circ}$ as shown in Fig.~\ref{fig:beta}). 
An edge-on configuration would have been more favorable to the orbit reconstruction process since the information contained in the Doppler measurements would have been stronger, likely leading to slightly flatter/smaller residuals. 

For Doppler data, an important step is to correct the propagation delay for tropospheric perturbations. This correction is performed using the VMF1 model \citep{Boehm_2006}.

\subsubsection{Landmarks}

Rosetta was equipped with two scientific cameras, the WAC (Wide Angle Camera) and the NAC (Narrow Angle Camera). They are both part of the OSIRIS (Optical, Spectroscopic, and Infrared Remote Imaging System) instrument \citep{Keller_2007}, which acquired about 100,000 images during the whole mission. 

\begin{figure*}[ht]

  \begin{minipage}[c]{.58\textwidth}
	\centering
	\includegraphics[width=\textwidth]{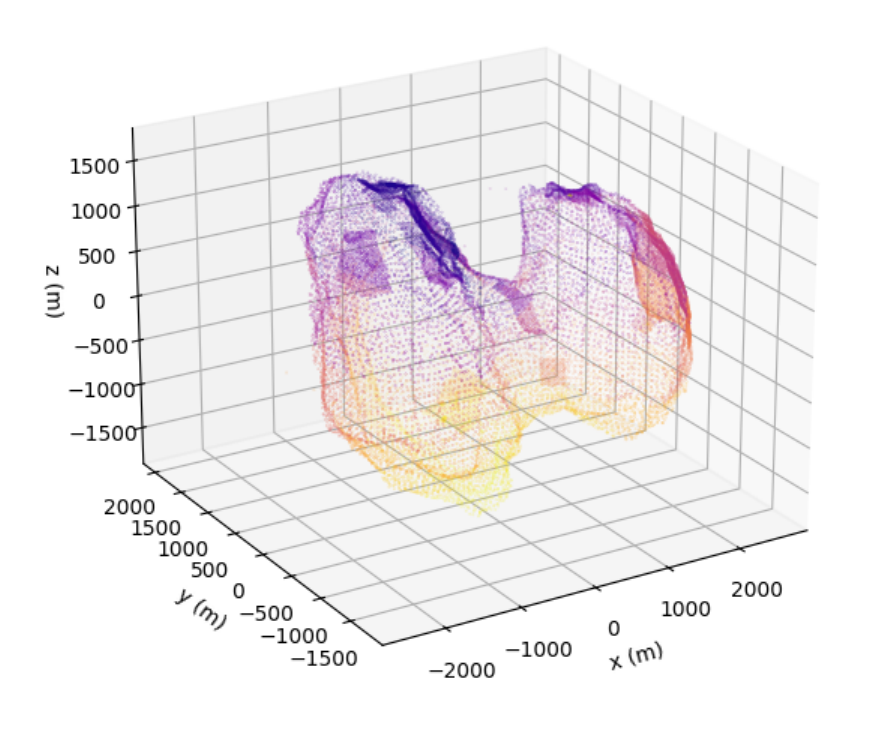}
  \end{minipage}%
  \begin{minipage}[c]{.42\textwidth}
    \includegraphics[width=\textwidth]{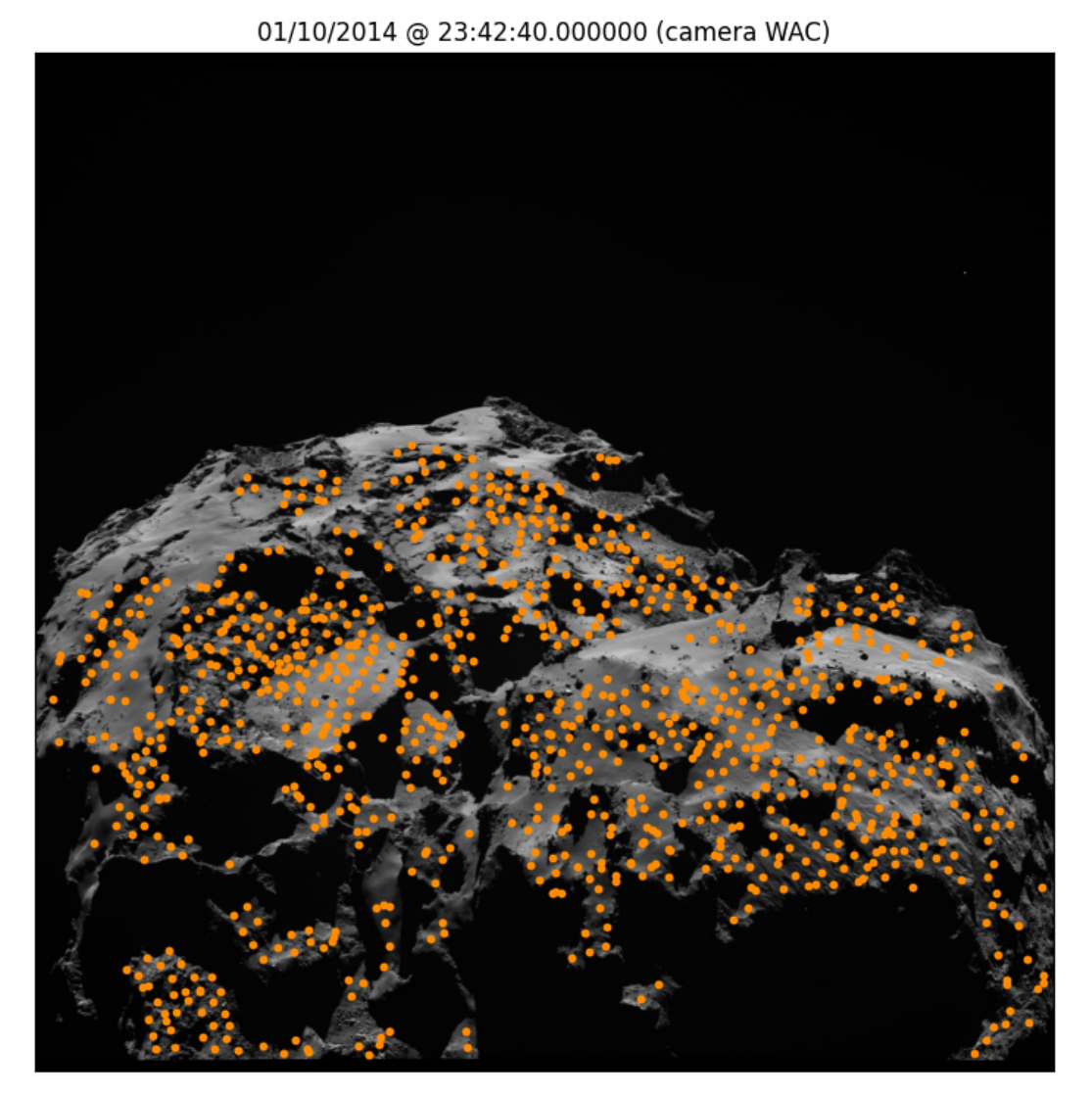}
  \end{minipage}
    \caption{(left) Position of the landmarks in the comet's body fixed frame. The total number of landmarks used in this study is 25570. (right) Landmarks identification over a picture, here with 968 visible landmarks}
    \label{fig:landmarks}
\end{figure*}

A subset of about 49,000 OSIRIS images of the nucleus have been analyzed with the SPC software  \citep{Gaskell_2008,Jorda_2016}. Among other products, the method provides landmarks coordinates in the body-fixed frame (see Fig.~\ref{fig:landmarks} left panel). As described in section~\ref{sec:shape}, these landmarks correspond to the center of small squared maplets in the SPC method.
The pixel coordinates of these landmarks are calculated from a stereo analysis and saved for each image having an intersection with the corresponding maplet.
An example of such landmark coordinates in a given image is shown in Fig.~\ref{fig:landmarks} right panel. Since we also know their coordinates in the body-fixed reference frame, we can obtain information about the position of the spacecraft with respect to the nucleus at the time of acquisition of the images. 
Finally, the stereo analysis also provides the orientation of the camera in the body-fixed frame for each OSIRIS image registered in SPC.

Although the SPC approach implements accurate distortion corrections, the archived SPC landmark coordinates refer to ``level 1'' images, which are not corrected from the optical distortion of the two cameras (NAC and WAC). It is therefore necessary to reprocess them with the model based on the in-flight geometric calibration of the cameras. The distortion is rather high, especially on the WAC, with displacements of tens of pixels at the edges. This correction is achieved by fitting 4th-order polynomial models whose coefficients are estimated by astrometric calibration of starfields during the commissioning phase of the mission.

Since the position of the spacecraft is estimated at each iteration, it is necessary to also adjust its attitude relative to the comet. To achieve this, the camera frame is rotated to fit each picture using three estimated angles. Such adjustment of the camera's pointing also permits the correction of $\pm$~30~arcsec errors relative to the commanded attitude of the spacecraft, which are due to thermo-elastic deformations of the spacecraft structure. The amplitude of these errors is quantified beforehand by comparing the star tracker quaternion with the pointing direction deduced from an astrometric analysis of OSIRIS starfield images acquired during the commissioning phase.

The attitude is independently re-estimated for each image, from the image itself. Then small rotations for each image were estimated. This allows to correct small errors in the estimated distortion corrections, thus reducing the level of noise.

\section{Methodology}
\label{sec:method}

To make use of these different types of data, a dynamical model is configured in the GINS software. The parameters of the model, primarily the gravity field coefficients, are then adjusted to fit the observations. The process is described in more details hereafter.

\subsection{Preliminary gravity sensitivity study} \label{grav_sensi}
\begin{figure*}[ht]
     \centering
     \includegraphics[width=0.48\textwidth]{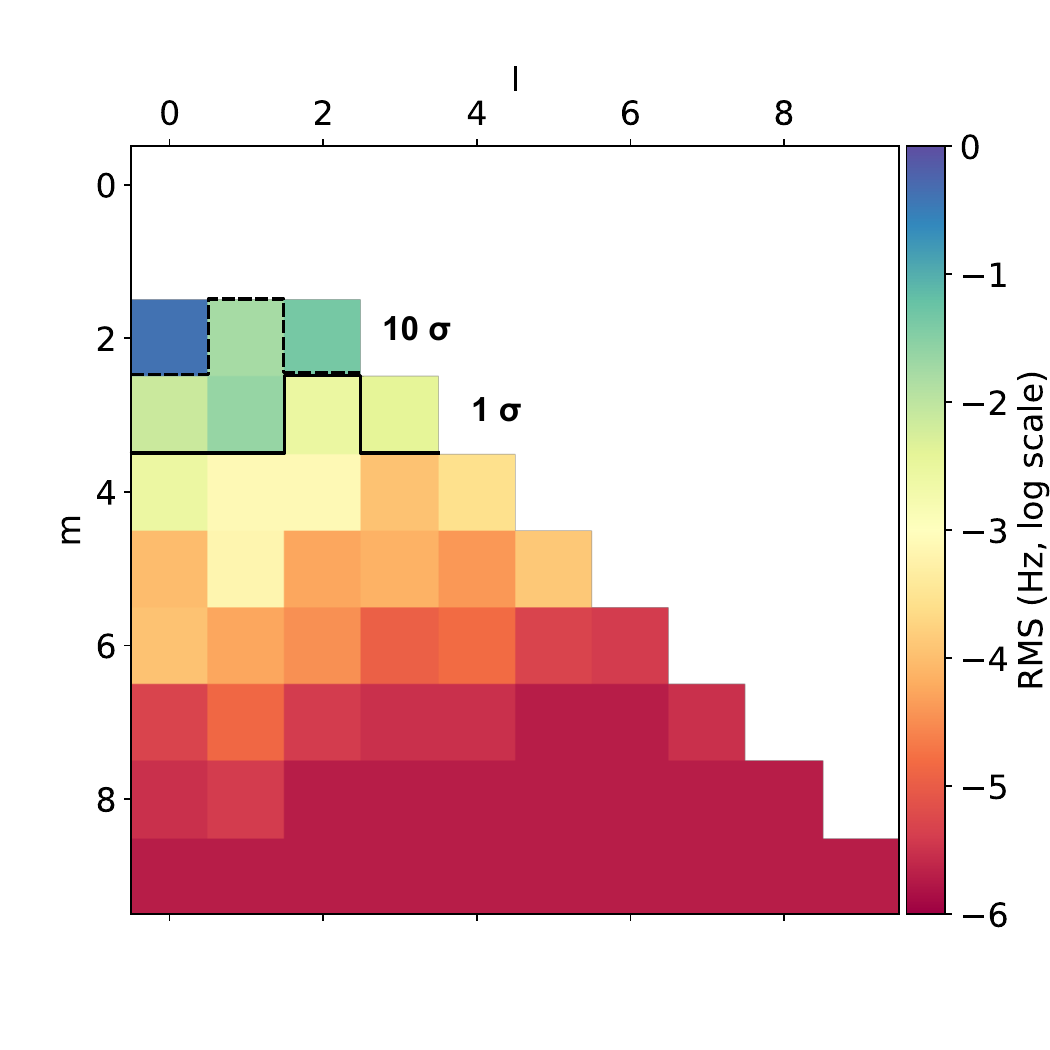}
     \includegraphics[width=0.48\textwidth]{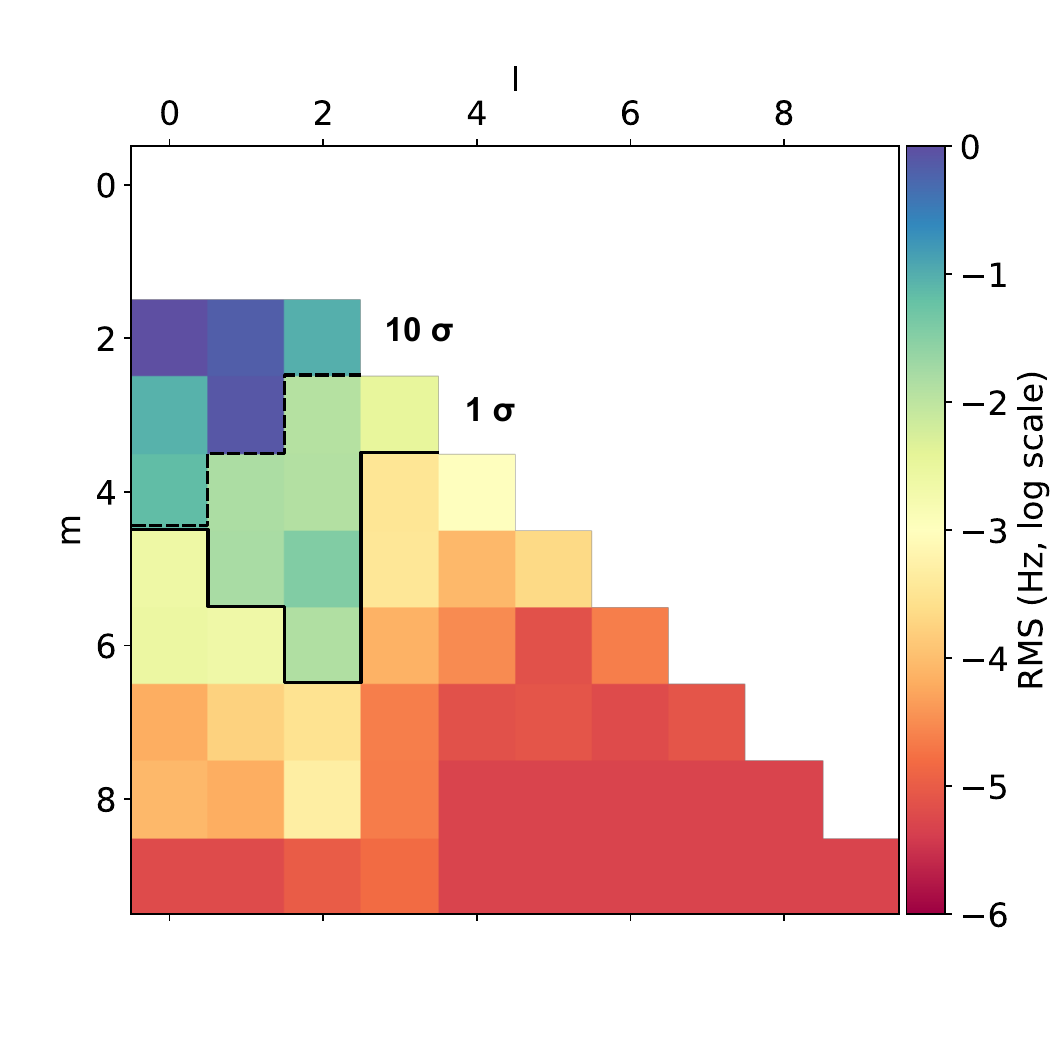}
     \caption{Theoretical Doppler shift induced by each (separate) spherical harmonic coefficient (only $\bar{C}_{lm}$ are shown here, $\bar{S}_{lm}$ have similar behaviour) for the longest and closest arc before (left) and after (right) perihelion. The dashed and solid black lines are boundaries delimiting coefficients with an impact of ten times the Doppler noise ($10\sigma$) and of the order of the Doppler noise ($1\sigma$), respectively.}
     \label{fig:sensi_1416}
\end{figure*}

Before estimating the gravity field coefficients, we need to determine which ones  actually have an observable signature in the data. Estimating non-observable (or weakly observable) parameters will result in inaccurate estimates, possibly leading to an overall unrealistic (i.e. non-physical) solution. We thus perform a sensitivity study to identify the highest spherical harmonic degree and order that can be determined by Rosetta. For that, we apply the following procedure:
\begin{enumerate}
    \item[1.] The longest arc with closest distance to the comet (i.e. Arc N=09 in Tab.~\ref{tab:arcs_def_14} and Arc N=08 in Tab.~\ref{tab:arcs_def_16}) is approximated at best using a fully keplerian orbit (GM only).
    \item[2.] The theoretical Doppler observations for this case are simulated using GINS.
    \item[3.] Then, the orbit is perturbed by switching a single gravity field coefficient at a time from 0 to its `homogeneous value'.
    \item[4.] The resulting Doppler observations are again simulated and compared to the unperturbed ones.
\end{enumerate}
If the deviation is higher than the noise measurement, the coefficient is considered observable. If it is comparable or below, the coefficient will be either weakly observable or not observable at all, and should not be estimated. \\
Results of the sensitivity analysis are shown in Fig.~\ref{fig:sensi_1416} for the $\bar{C}_{lm}$. Results for the $\bar{S}_{lm}$ are basically the same. 
Attention must be paid on different approximations made for this sensitivity study. First, only Doppler observations were considered, which explains why degree-1 is ignored here. Adding landmarks will improve the solution and allow determining degree-1 coefficients. 
Second, the gravity field used in this sensitivity study is based on the assumption of homogeneous mass distribution. If some coefficients turn out to be weaker or stronger than their homogeneous counterpart, the results of the sensitivity study could change a little. These results must therefore be interpreted only in terms of orders of magnitude. 
Third, we quantify the influence of each gravity coefficient separately. The correlations are thus ignored and a linear combination of some of these coefficients could actually leave the Doppler signal unchanged. These results should therefore be considered as an achievable maximum.

\noindent
Based on this study, we have decided to define two calculation cases: 
\begin{itemize}
    \item {\bf ``Case 0/2" :} in this first case we estimate the coefficients $\bar{C}_{10}$, $\bar{C}_{11}$, $\bar{S}_{11}$, $\bar{C}_{20}$, $\bar{C}_{22}$ and $\bar{S}_{22}$ over the entire mission (global parameters), and we estimate $\bar{C}_{00}$ (i.e. the GM) before and after perihelion (period-dependent parameter).
    \item {\bf ``Case 2/4" :} in the second case we estimate the coefficients $\bar{C}_{21}$, $\bar{S}_{21}$, $\bar{C}_{30}$, $\bar{C}_{31}$, $\bar{S}_{31}$ and $\bar{C}_{40}$ over the whole mission (global parameters), while $\bar{C}_{00}$, $\bar{C}_{10}$, $\bar{C}_{11}$, $\bar{S}_{11}$, $\bar{C}_{20}$, $\bar{C}_{22}$ and $\bar{S}_{22}$ are estimated separately either from data before perihelion or from those acquired after perihelion.
\end{itemize} 

\subsection{Measurements editing and arcs selection} \label{subsec:data_qual}
Doppler measurements are affected by white noise of the order of a few milli-hertz. For some reasons (technical or operational), some data points are doubtful. An appropriate editing method would allow us to reject these bad or questionable observations, but it is not straightforward to decide which data points are ``good'' or not. In GINS, data editing is done in two steps. After comparison with the pre-fit orbits of Rosetta, residuals with a root mean square (RMS) larger than 5Hz are clear outliers and discarded (This represents an average of less than $2\%$ of measurements over the selected arcs). Additionally, measurements acquired when the spacecraft elevation as seen from the ground station is below the empirical threshold of $12 \deg$ are also discarded, since the accuracy of the tropospheric correction decreases with this angle. 
The orbit is determined in an iterative process, in which dynamical parameters are fitted to the observations. At each iteration, GINS filters out additional data points with residuals larger than $3 \sigma$ of the residuals RMS. After the orbit has converged (i.e. the post-fit residuals RMS is stable), the entire arc is kept or not depending on the percentage of rejected observations (too much data eliminated is not acceptable because the orbit may then converge to an inaccurate solution). Only a handful of arcs (5 arcs before perihelion and 3 arcs after perihelion in total) are manually eliminated based on this criterion.\\
A second set of arcs has been rejected due to unexplained jumps in the Doppler signal of the order of $5~mHz$ at exactly midnight recording time. Since there is no way to know which data are correct (before or after the discontinuity), the whole arc is manually discarded when such a phenomena occurs. This leads to the rejection of 2  arcs pre-perihelion and 2 post-perihelion\\
A third set of 12 arcs between June 1st and July 20th, 2016, have all been rejected because of the numerous bad measurements identified during that period, which is thought to be essentially linked to the issue described above.\\
Finally, 16 arcs (15 pre- and 1 post-perihelion) with minimum distance between the spacecraft and the comet greater than 25 km have been rejected due to low sensitivity to the gravoity field.\\

Tab.~\ref{tab:sum_data} summarises all the measurements a priori available and those actually used in the calculation for the two settings considered in this study, namely 'Case 0/2' and the 'Case 2/4' as defined above.

\begin{table*}[ht]
    \caption{Summary of the Doppler tracking data and landmark measurements used for 67P/C-G gravity solutions depending on the estimated parameter setup used.}
    \label{tab:sum_data}
    \begin{tabular}{lc|rr|rr|rr}
    \toprule
        ~               &         & \multicolumn{2}{c|}{Total} & \multicolumn{2}{c|}{Case 0/2}  & \multicolumn{2}{c}{Case 2/4} \\
        ~               & Number  & Number of     & Number of  & Number of     & Number of  & Number of     & Number of \\
        ~               & of arcs & 2-way Doppler & Landmarks  & 2-way Doppler & Landmarks  & 2-way Doppler & Landmarks \\ \midrule
        Pre-perihelion  & 15      & 246 874        & 474 822   & 238 510       & 455 349    & 238 477       & 455 717 \\
        Post-perihelion & 43      & 242 064        & 659 738   & 235 652       & 631 426    & 235 696       & 633 721 \\ 
        Full            & 58      & 488 938        & 1 134 560 & 474 162       & 1 086 775  & 474 173       & 1 089 438 \\\bottomrule
    \end{tabular}
\end{table*}

\subsection{Parameters estimation procedure}
\subsubsection{Estimated Parameters setup} 
 We now estimate the parameters of our model by fitting them to observations using a least squares inverse approach. The model has been described in the previous sections. The 13.8k estimated parameters, included the target gravity  coefficients,  are grouped in Tab.~\ref{tab:param}.

\begin{table*}
\centering
\caption{Free parameters of the model. The arc-dependent parameters (solved in GINS) are denoted AD, the global parameters (solved by accumulating the normal equations for each arc) are denoted GP, and period-dependent parameters, which are estimated using data collected either before or after perihelion, are denoted PD.} \label{tab:param}
\begin{tabular}{lcc}
\toprule
{\bf Parameter Name} & Type of parameter & Total number\\ 
\midrule
{\bf Initial Spacecraft position and velocity} & AD & 348 (58 $\times$ 6)\\
{\bf Wheel of Loading (WoL) residual $\Delta V$} & AD & 690 (230 $\times$ 3)\\
{\bf Camera orientation} & AD & 12774 (4258 $\times$ 3) \\ 
{\bf 67P/C-G gravity field} & PD & 6, 8 or 16 \\
{\bf SRP Scale factor} & GP & 1 \\
\bottomrule
\end{tabular}
\end{table*}

Spacecraft position and velocity components at the start of each arc are estimated. To these six initial state parameters, three additional ones are added for each of the 230 desaturations of the inertial wheels (one velocity increment per axis), to compensate for the  small residual $\Delta V$ that they induce. The camera attitude is adjusted using three angles $(\alpha_X, \alpha_Y, \alpha_Z)$ representing rotations around the X, Y and Z camera axes, respectively. We introduce a weak constraint to limit the amplitude of the corrections on these angles. 
The gravitational attraction of the comet is classically modeled using a spherical harmonics expansion as described above. The strategy of estimation of the $\bar{C}_{lm},\bar{S}_{lm}$ of the comet follows the definition of the cases in section~\ref{grav_sensi}. Un-estimated coefficients up to degree and order 20 are still introduced into our gravity field model (see Supplement material), but they remain fixed to the values given by the shape model under the assumption of a uniform mass distribution (to stay as close as possible to real flight conditions).
Finally, one scaling factor is estimated over the entire mission to calibrate the solar radiation pressure force and account for the uncertainties of the macro-model of the spacecraft (i.e. the bus and solar panels models).

\subsubsection{Gravity field estimation procedure} 

As mentioned in section \ref{sec:modeling}, some of the forces taken into account depend on parameters that are either estimated (e.g. SRP and Stokes coefficients up to degree 2 or 4) or fixed to their a priori/model value (e.g. out-gassing and third-body effects). The problem is solved using the GINS software developed at CNES \citep{doc_algo_GINS}. GINS can propagate orbits and adjust parameters such as the gravity field to fit measurements at best. Because GINS can only adjust parameters arc per arc, it is combined with DYNAMO for multi-arc processing. DYNAMO is a tool belonging to the GINS software package which allows us to stack and solve systems of linear equations. The sequence of operations performed with the GINS/DYNAMO chain is as follows:
\begin{itemize}
    \item GINS is first used to get an initial estimate of AD parameters while the GP and PD parameters (as the gravity field coefficients) are kept fixed. Using an iterative least squares procedure, the initial position and velocity of the spacecraft are typically estimated at this stage of the POD process, along with other local parameters like WoL and cameras pointing. 
    \item Once convergence has been achieved, GINS produces the normal equations including all parameters, i.e. AD, PD and GP like the gravity field and the SRP scale factor.
    \item These equations are then combined using a specific weighting based on Helmert’s method \citep{Sahin1992} and solved with a truncated SVD method \citep{Hansen1987}.
    \item Then, the new PD and GP parameters (ie. the new gravity field and the new SRP scale factor) are injected back into GINS and a new "macro-iteration" (i.e. iteration of the complete GINS/DYNAMO chain) begins. In this case, several are necessary since the domain is not linear.
\end{itemize}

\noindent
For the orbit initialization, we extract a priori value of the spacecraft states at the beginning of each arc from the SPICE kernel (see Tab.~\ref{tab:kernels}). The gravity field is initialized using the shape model of the comet, computed using GILA software developed by \citep{Caldiero2024} assuming homogeneous mass distribution inside the comet. Therefore if a homogeneous-density solution exists, we start nearby. 
The so-obtained $\bar{C}_{lm},\bar{S}_{lm}$ up to degree and order 20 are reported in the Supplement material. The other estimated parameters are initialized at zero for WoL $\Delta V$ and the SRP scale factor is set to $F_S=1$ (see Eq.~\ref{eq:SRP}).\\

\noindent
The problem is non-linear due to the odd shape of the comet and the relative scarcity of the measurements, and some weak constraints are added to keep the solution within reasonable limits. In GINS and DYNAMO, constraints are quadratic penalties in the solution space that guide solutions whose parameter values are too far from their expected values (as opposed to hard constraints that would completely prohibit certain parameter values). The following constraints are imposed: the SRP scale factor is $1 \pm 1$, the WoL residual $\Delta V$ are $0 \pm 10^{-5} m/s$ and the camera attitude adjustments are $0 \pm 0.1 \deg$. No constraints are imposed on orbital and gravitational parameters.

As mentioned above, a regularisation of the inversion of the normal equation is applied using the singular value truncation method \citep{Hansen1987}. The threshold value selected is: $10^7$.

\section{Results}
\label{sec:results}

This section gathers the numerical results of the two cases defined above: "Case 0/2" and "Case 2/4".
In the former, the degree 1 and 2 coefficients are global parameters adjusted over all the arcs, while degree 0 (i.e. GM) is a period-dependent parameter estimated once using the pre-perihelion data only and a second time using the post-perihelion data only. In the  Case 2/4, degree 0, 1 and (partly) 2 are estimated per period, while the others are adjusted over the full set of data (PD and GP are separated using  the 10$\sigma$ boundary criterion shown in Figs.~\ref{fig:sensi_1416}).

\subsection{Adjustment of non gravitationnal parameters}

\begin{figure}[ht]
     \centering
     \includegraphics[width=\userfigwidth]{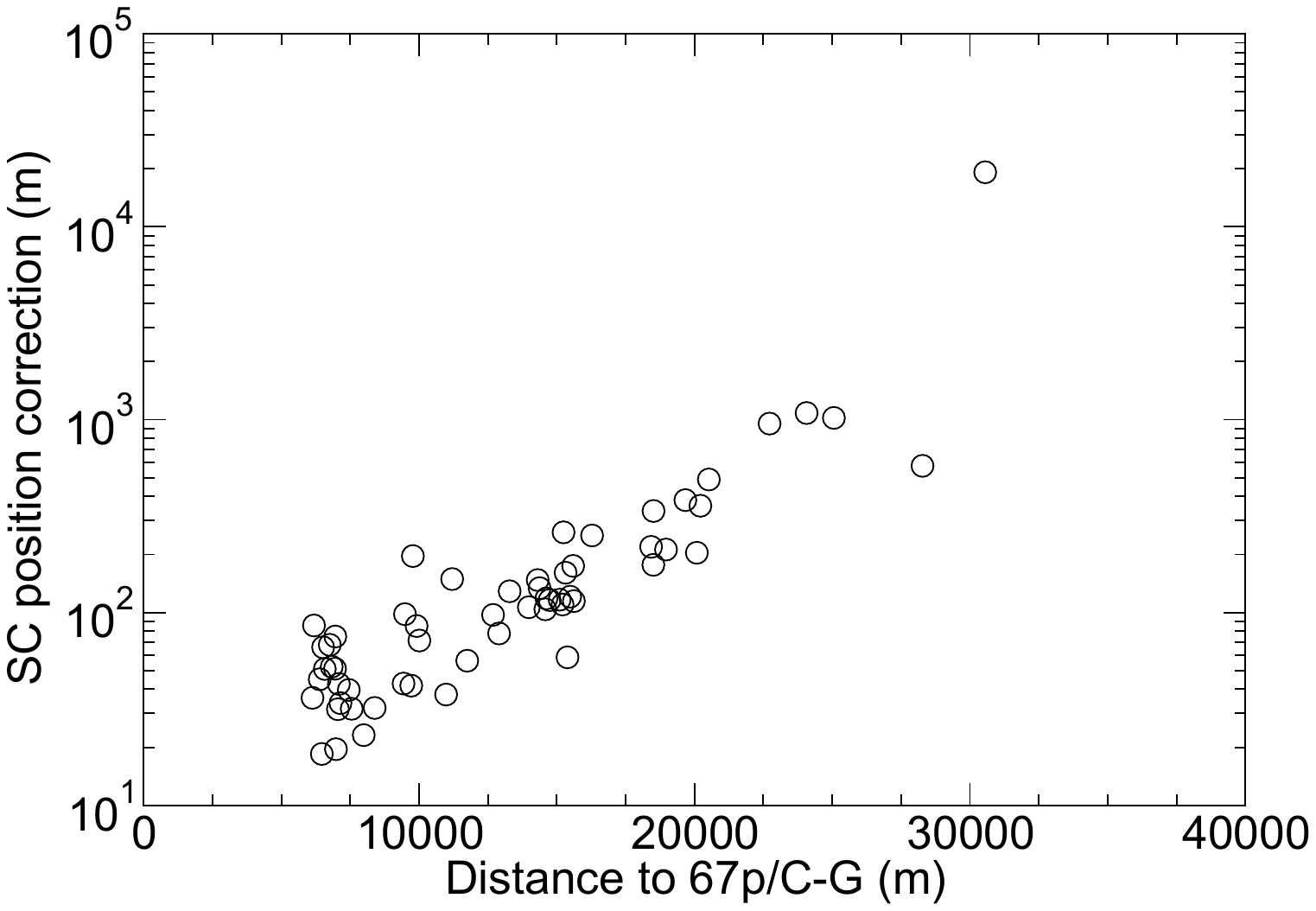}
     \caption{Corrections to Rosetta's initial position for each of the 58 arcs of this study as a function of the average distance from the spacecraft to the centre of the comet.}
     \label{fig:Dist_corr}
\end{figure}

On each arc, the position and initial velocity of the spacecraft are adjusted, along with the orientation of the camera for each photo taken and the residual $\Delta V$ for each wheel of loading manoeuvre. The adjustment of each of these parameters tells us about the relevance of our initial hypothesis/models. \\
We observe a significant correlation between the corrections of the initial states of Rosetta and the distance to the comet (see Fig.~\ref{fig:Dist_corr}). The closer the trajectory of the arc is to the center of the comet, the smaller the corrections in the initial position. This can be explained by the fact that our a priori values for the spacecraft initial position vector are taken from the SPICE kernel, provided by the ESOC navigation team, which included, just like us, both radiometric data and images (but coming from the navigation camera instead of OSIRIS like us) as described by \cite{Godard2016MultiarcOD}. As a result, both the orbit of ESOC/SPICE and ours should be comparable at short distance to the comet where images provide more accurate position of Rosetta with respect to the comet. \\

The 230 $\Delta V$ residuals of the adjusted WoLs for all arcs have a mean of zero, with a standard deviation of 0.4 mm/s. This is one order of magnitude below the noise, which gives us confidence in our solution. High $\Delta V$ estimates would have indicated a limitation of our dynamical model where neglected phenomena could have been wrongly absorbed by these parameters.\\
The corrections in the orientation of the OSIRIS cameras are on average a few tenths of a degree, with a standard deviation of 0.4 degrees. This is of same order of magnitude as the  accuracy of the comet orientation model (see Fig.~\ref{fig:AD_CG67P} and associated discussions in Sec.~\ref{ssec:ephrotor}), which again gives us confidence in our fits.

At the end of all the iterations, the SRP scaling factor is estimated to be $F_S=1.000052$ with a standard deviation of $4.6 \times 10^{-6}$. The fact that $F_S$ is very close to 1 and that its standard deviation is very low tells us that the force is well observed and determined and that one scaling factor as a global parameter (i.e. used to fit solar pressure over the full set of measurements) is well suited for Rosetta.

\subsection{Convergence of gravity coefficients}
\label{res:conv}


The results for our degraded case (case 0/2) are obtained in a single step of 12 iterations where the coefficient $\bar{C}_{00}$ is estimated before and after perihelion and $\bar{C}_{10}$, $\bar{C}_{11}$, $\bar{S}_{11}$, $\bar{C}_{20}$, $\bar{C}_{22}$ and $\bar{S}_{22}$ are estimated over the two periods combined. It should be noted that all other coefficients are left at the values deduced from the shape.\\
\noindent
The results for our nominal case (case 2/4) are obtained in two steps: for the first 12 iterations, coefficients $\bar{C}_{00}$, $\bar{C}_{10}$, $\bar{C}_{11}$, $\bar{S}_{11}$, $\bar{C}_{20}$, $\bar{C}_{22}$ and $\bar{S}_{22}$ are estimated before and after the perihelion separately and $\bar{C}_{21}$ and $\bar{S}_{21}$ are estimated on both period combined. For 12 additional iterations, coefficients $\bar{C}_{21}$ and $\bar{S}_{21}$, $\bar{C}_{30}$, $\bar{C}_{31}$, $\bar{S}_{31}$ and $\bar{C}_{40}$ are estimated for both periods combined, and the previously estimated coefficients are left free, but constrained to the value obtained at the twelfth iteration with a weight of 3$\sigma$.\\

\begin{figure*}[ht]
\begin{minipage}[c]{.6\textwidth}
 \centering
 \includegraphics[width=\textwidth]{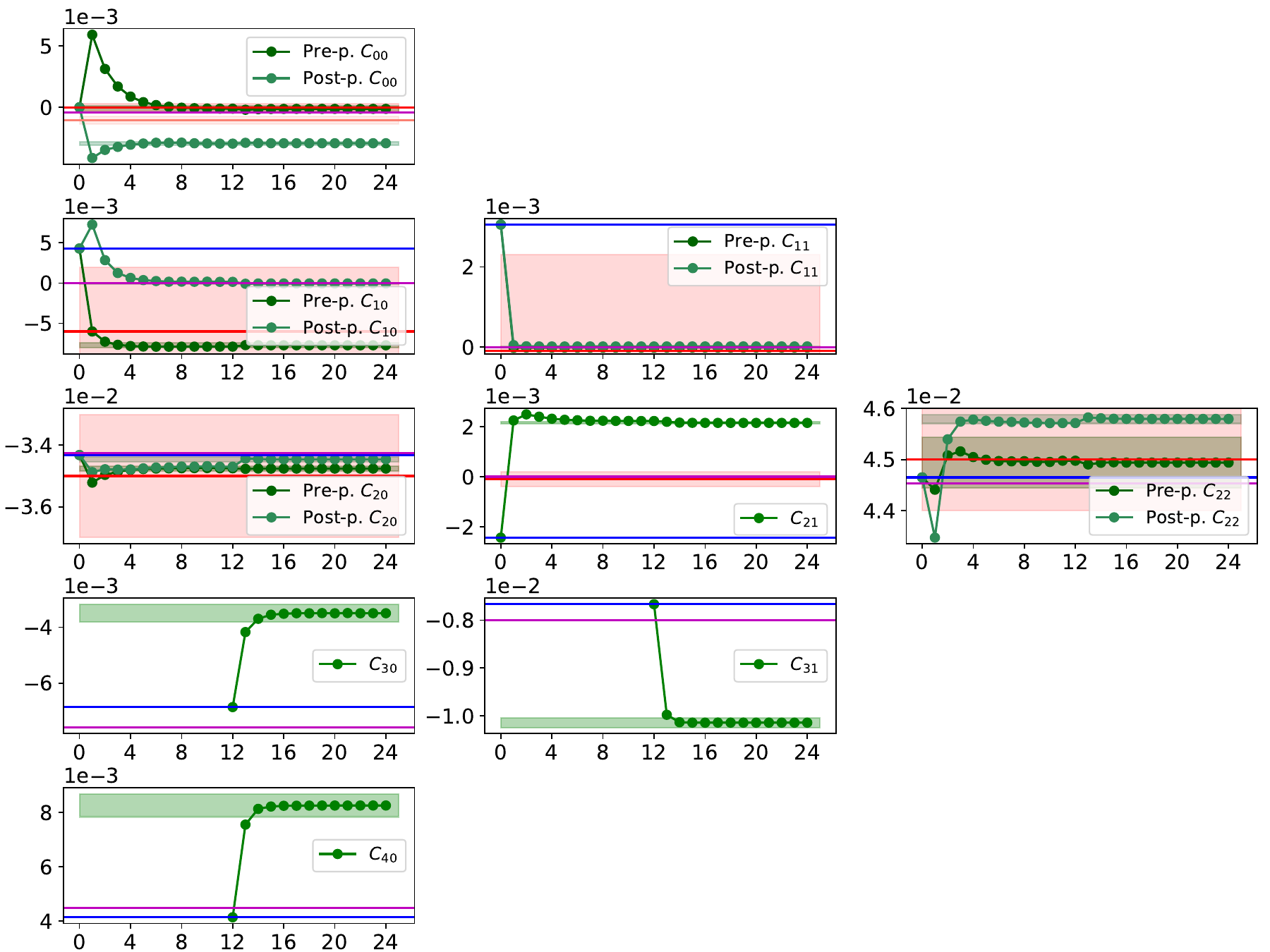}
\end{minipage}%
\begin{minipage}[c]{.4\textwidth}
 \centering
 \includegraphics[width=\textwidth]{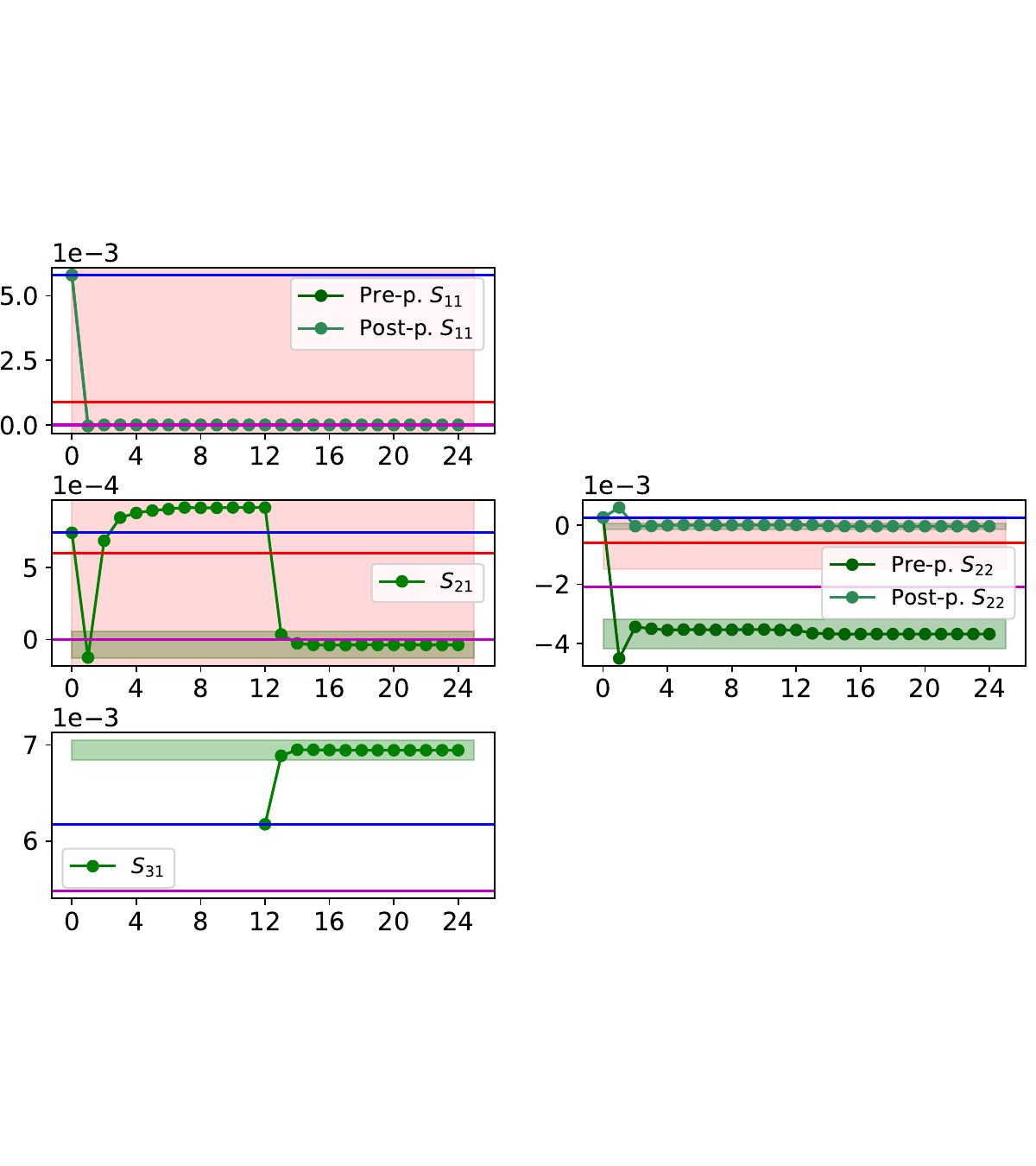}
\end{minipage}

\caption{Estimated coefficients $\bar{C}_{lm}$ and $\bar{S}_{lm}$ over iterations. Blue lines are the ``shape coefficients'', red lines are  solutions from \cite{Patzold_2019} and purple line are solutions from \cite{Godard2016MultiarcOD}. Red shaded areas are formal uncertainties (i.e. $1\sigma$) of \cite{Patzold_2019}, while green shaded areas are ten times the standard deviations (i.e. $10\sigma$)  of our convergence calculation. Top left panel shows in fact $\bar{C}_{00}-1$ solutions for easier reading.}
\label{fig:conv_CS}

\end{figure*}

\noindent
Fig.~\ref{fig:conv_CS} shows the evolution of our `Case 2/4' estimates of $\bar{C}_{lm},\bar{S}_{lm}$ over the iterations. The convergence profile for the `Case 0/2' is quite similar to that of the `Case 2/4'. As shown on the figure, all the parameters converged well, with a clear distinction between both periods estimates of $\bar{C}_{00}$ and $\bar{C}_{10}$. The coefficients $\bar{C}_{20}$ before and after perihelion both converge to the same value, unlike the $\bar{C}_{22}$ and $\bar{S}_{22}$ coefficients which differ by $\sim1\cdot10^{-3}$ (i.e. $\sim20\sigma$) and $\sim3.5\cdot10^{-3}$ (i.e. $\sim70\sigma$) respectively between the pre- and post-perihelion values. The coefficients $\bar{C}_{21}$ and $\bar{S}_{21}$ are released from the first step, but are not distinguished between the pre-perihelion and post-perihelion phases of the mission. While $\bar{C}_{21}$ converges very quickly to a value that it maintains in the second step, $\bar{S}_{21}$ estimates undergoes a jump between the two steps of the algorithm. This is the only parameter that seems to be significantly affected by our two-step adjustment strategy. However one needs to weigh that observation up, because the amplitude of $\bar{S}_{21}$ is at least an order of magnitude lower than the others.

\subsection{State of the art comparison}
Our new solution of 67P/C-G gravity field (Case 2/4) is compared to previous values in Tabs.~\ref{tab:grav_field_01}, \ref{tab:grav_field_2} and \ref{tab:grav_field_deg_sup}. From our estimate of $\bar{C}_{00}$ using data before (Pre-P.) and after (Post-P.) perihelion, we get the following values for the mass of the comet:
\begin{align*}
    M_\text{Pre-P.}  &= (9.980 \pm 0.00025) \times 10^{12} kg \\
    M_\text{Post-P.} &= (9.952 \pm 0.00014) \times 10^{12} kg \\
    \Delta M &= (28.00 \pm 0.29) \times 10^9 kg
\end{align*}
$\Delta M$ is the mass lost during the perihelion pass. It represents about 0.28\% of the total mass of the comet. This is more than two times larger than previous estimates of 0.1\% \citep{Patzold_2016,Patzold_2019}. \\

\noindent
Our (classical) measurement modeling assumes that all the data points are affected by a decorrelated error, but this approximation is almost certainly wrong. In fact, Doppler measurements are all derived from the same electronic process and may be affected by coloured noise due to auto-correlation of the signal. As for the landmark measurements, batches of several hundred of them are identified per image (see Fig.~\ref{fig:landmarks}), making them subject to a common degree of error. This simplification of the measurement modelling therefore leads to an underestimation of the formal standard deviations obtained after solving the least-squares problem. There is no way, except through a complicated simulation, to estimate these realistic correlations. The sigmas presented here are therefore very optimistic, and should be interpreted with great caution. A possible heuristic would be to multiply all the standard deviations by ten as done for Fig.~\ref{fig:conv_CS}, but we prefer to provide the raw GINS/DYNAMO outputs in the tables of this section, although the central values are truncated to significant figures (i.e. $\pm10$ times the sigmas).

\begin{table*}[ht]
\caption{Estimated GM (degree 0) and degree 1 coefficients of 67P/C-G using Doppler and landmark data (D\&L) for `Case 0/2' and `Case 2/4'. For comparison, the values obtained under the assumption of Uniform Mass Distribution (UMD) and the previous results from \cite{Patzold_2019} using Doppler (D) and from \cite{Godard2016MultiarcOD} and \cite{GAO2023} using D\&L are also reported. All uncertainties are formal errors ($1\sigma$). }\label{tab:grav_field_01}

\begin{tabular*}{\textwidth}{@{}C|c|c|ccc@{}}
\toprule
       & \multirow{ 2}{*}{Period}  & GM   & \multirow{ 2}{*}{$\bar{C}_{10}$}  & \multirow{ 2}{*}{$\bar{C}_{11}$}  & \multirow{ 2}{*}{$\bar{S}_{11}$}\\
       &                           & $[m^3/s^2]$  & & & \\
\midrule
        Shape (UMD)   & - & & $0.0043$ &  $0.0031$ & $-0.0058$  \\
\midrule
   \multirow{ 2}{*}{Pätzold (D)} & Pre-P. & $666.2 \pm 0.2$ & \multirow{ 2}{*}{$-0.006 \pm 8.1\,10^{-3}$}  & \multirow{ 2}{*}{$-0.0001 \pm 2.4\,10^{-3}$} &  \multirow{ 2}{*}{$0.0009 \pm 6.1\,10^{-3}$}  \\
           & Post-P. & $665.5 \pm 0.1$   \\
\midrule
  Godard (D\&L)  & Full & $665.9 \pm 0.3$ & $0.0013 \pm 1.5\,10^{-4}$ & $0.0000$ &  $0.0000$ \\
\midrule
  Gao (D\&L)  & Pre-P. & $665.3 \pm 0.1$ & $-0.0013 \pm 3.0\,10^{-4}$ & $0.0000 \pm 1.0\,10^{-4}$ &  $0.0018 \pm 1.0\,10^{-4}$ \\
\midrule
  \multirow{ 2}{*}{Case 0/2 (D\&L)} 
  & Pre-P.  & $666.4 \pm 0.03$  & \multirow{ 2}{*}{$-0.0016 \pm 2\,10^{-5}$}   & \multirow{ 2}{*}{$0.0000\pm 3\,10^{-5}$} &  \multirow{ 2}{*}{$0.0000 \pm 3\,10^{-5}$}  \\
  & Post-P. & $664.3 \pm 0.02$  \\
\midrule
  \multirow{ 2}{*}{Case 2/4 (D\&L)} 
  & Pre-P.  & $666.1 \pm 0.02$ & $-0.0077 \pm 3\,10^{-5}$ & $-1.18\,10^{-5} \pm 3\,10^{-7}$ & $-2.68\,10^{-6} \pm 3\,10^{-7}$   \\
  & Post-P. & $664.2 \pm 0.01$ &  $-4.51\,10^{-5} \pm 1\,10^{-5}$ & $ 1.63\,10^{-5} \pm 2\,10^{-7}$ & $ 7.34\,10^{-6} \pm 2\,10^{-7}$ \\
\bottomrule
\end{tabular*}
\end{table*}

\begin{table*}[ht]
\caption{Estimated degree 2 coefficients of 67P/C-G using Doppler and landmark data (D\&L) for `Case 0/2' and `Case 2/4'. For comparison, the values obtained under the assumption of Uniform Mass Distribution (UMD) and the previous results from \cite{Patzold_2019} using Doppler (D) and from \cite{Godard2016MultiarcOD} and \cite{GAO2023} using D\&L are also reported. All uncertainties are formal errors ($1\sigma$). }\label{tab:grav_field_2}

\footnotesize
\begin{tabular*}{\textwidth}{@{}C|c|ccccc@{}}
\toprule
                                   & Period   & $\bar{C}_{20}$ & $\bar{C}_{21}$ & $\bar{S}_{21}$ & $\bar{C}_{22}$ & $\bar{S}_{22}$ \\
\midrule
 Shape (UMD)                       & -    & $-0.0343$ & $-0.0024$ & $0.0007$ & $0.0445$ & $-0.0001$ \\
\midrule
 Pätzold (D)                       & Full    & $-0.035 \pm 2.0\,10^{-3}$  & $-0.0001 \pm 3.0\,10^{-4}$ & $0.0006 \pm 8.0\,10^{-4}$ & $0.045 \pm 1.0\,10^{-3}$  & $-0.0006 \pm 9.0\,10^{-4}$ \\
\midrule
 Godard (D\&L)                     & Full    & $-0.0343 \pm 3.1\,10^{-4}$ & $0.0000$ & $0.0000$ & $0.0445 \pm 1.3\,10^{-4}$ & $-0.0021 \pm 1.3\,10^{-4}$ \\
 \midrule
 Gao (D\&L)                     & Pre-P.    & $-0.0351 \pm 1.6\,10^{-4}$ & $0.0000$ & $0.0000$ & $0.0447 \pm 3.0\,10^{-4}$ & $-0.0022 \pm 2.8\,10^{-4}$ \\
 \midrule
 Case 0/2 (D\&L)                   & Full    & $-0.0347 \pm 1\,10^{-5}$  & $-0.0024$ & $0.0007$ & $0.0455 \pm 2\,10^{-5}$  & $-0.0001 \pm 2\,10^{-5}$ \\
\midrule
 \multirow{ 2}{*}{Case 2/4 (D\&L)} & Pre-P.  & $-0.0348 \pm 8\,10^{-6}$  &  \multirow{ 2}{*}{$0.0022 \pm 5\,10^{-6}$} & \multirow{ 2}{*}{$-3.6\,10^{-5} \pm 9\,10^{-6}$}  & $0.0449 \pm 5\,10^{-5}$ & $-0.0037 \pm 5\,10^{-5}$ \\
                                   & Post-P. & $-0.0345 \pm 8\,10^{-6}$  &                                               &                                                     & $0.0458 \pm 9\,10^{-6}$ & $-3.5\,10^{-5} \pm 1\,10^{-5}$ \\
\bottomrule
\end{tabular*}
\end{table*}
\noindent
As shown in Tab.~\ref{tab:grav_field_01} and Tab.~\ref{tab:grav_field_2}, our estimates are generally speaking in good agreement with the results of \cite{Patzold_2019}. Some discrepancies can nevertheless be observed in the solutions of $\bar{C}_{11}$, $\bar{S}_{11}$, $\bar{C}_{21}$ and $\bar{S}_{21}$, which can be partly explained by a different frame definition since newer SPICE kernels are used here. The reference frame is also strongly linked to the landmark definition.\\
Our higher degrees solutions are shown in Tab.~\ref{tab:grav_field_deg_sup}.  Not estimated by \cite{Patzold_2019}, we only compare them to those provided by \cite{Godard2016MultiarcOD}. Since the latter provided  non-normalised Stokes coefficients estimates, relative to a different reference radius of 1000 m, we  report here coefficients that have been converted using the same standard as the others, i.e. normalised and with a radius of 2650 m. Not provided by these authors, the coefficients $\bar{C}_{11}$, $\bar{S}_{11}$ and $\bar{C}_{21}$, $\bar{S}_{21}$ have been set to 0 for the ``Godard solution''. Note that  \cite{Godard2016MultiarcOD} did not distinguish between pre- and post-perihelion, so didn't estimate any loss of mass at perihelion. For these degree 3 and 4 coefficients, significant discrepancies are observed. For low degrees, the uniform distribution was indeed a legitimate conclusion given the accuracy of previous results. However the increased accuracy here due to the addition of landmarks shows some deviations from the previous conclusion. 

\begin{figure}[ht]
     \centering
     \includegraphics[width=\userfigwidth]{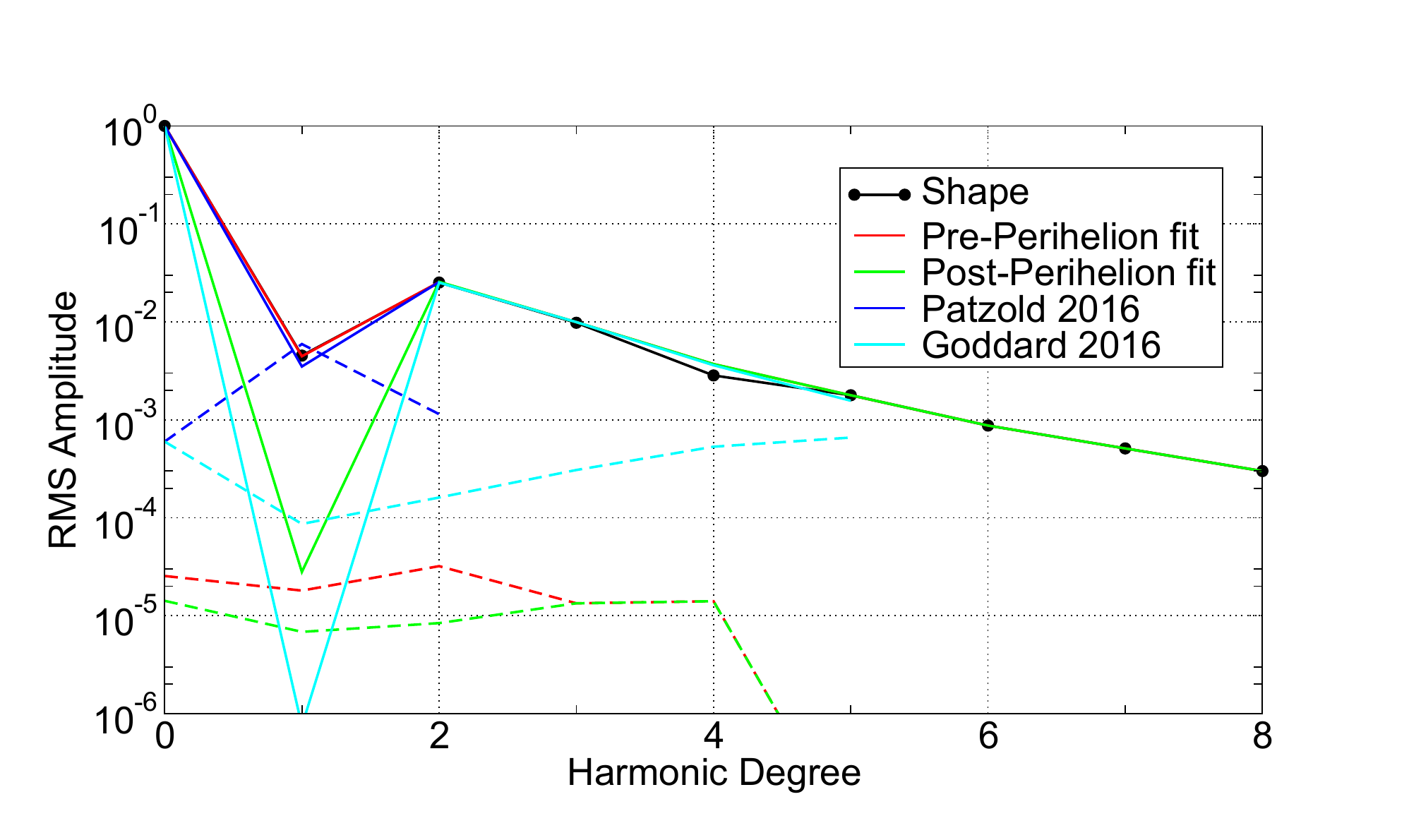}
     \caption{The solid lines are the powers spectrum of the solutions and the dashed lines are the powers spectrum of the standard deviations.}
     \label{fig:spectre}
\end{figure}

\begin{table*}[ht]
    \caption{Estimated degree 3 and 4 coefficients of 67P/C-G using Doppler and landmark data (D\&L) for `Case 0/2' and `Case 2/4'. For comparison, the values obtained under the assumption of Uniform Mass Distribution (UMD) and the previous results from \cite{Godard2016MultiarcOD} and \cite{GAO2023} using D\&L are also reported. All uncertainties are formal errors ($1\sigma$). }
    \label{tab:grav_field_deg_sup}
    \begin{tabular}{c|cccc}
    \toprule
                        & $\bar{C}_{30}$ & $\bar{C}_{31}$ & $\bar{S}_{31}$ & $\bar{C}_{40}$  \\ \midrule
        Shape (UMD)     & $-0.0067$  & $-0.0076$  & $0.0062$   & $0.0042$  \\ 
        Godard (D\&L)   & $-0.0076\pm 3.7\,10^{-4}$  & $-0.0080\pm 3.7\,10^{-4}$  & $0.0055\pm 3.4\,10^{-4}$   & $0.0045\pm 6.3\,10^{-4}$  \\
        Gao (D\&L)      & $-0.0064\pm 6.7\,10^{-4}$  & $-0.0101\pm 6.7\,10^{-4}$  & $0.0113\pm 7.4\,10^{-4}$   &   \\
        Case 2/4 (D\&L) & $-0.0035\pm 3\,10^{-5}$  & $-0.0101\pm 1\,10^{-5}$  & $0.0069\pm 1\,10^{-5}$   & $0.0083\pm 4\,10^{-5}$  \\  \bottomrule
    \end{tabular}
\end{table*}
\noindent
Fig.~\ref{fig:spectre} shows the power spectrum of our gravity solutions (computed with Eq.~\ref{rms:CS}) superimposed with that of the UMD model and of the solution of \cite{Patzold_2019}. The power spectrum of the standard deviations, also reported on that figure, were computed as follows:
\begin{equation}
\sigma_{P_l}=\sqrt{\frac{\sum_{m=0}^l \left(\sigma_{\bar{C}_{lm}}^2+\sigma_{\bar{S}_{lm}}^2\right)}{2l+1}}. \label{rms:sigma}
\end{equation}
\noindent
The significant improvement of our solution with respect to \cite{Patzold_2019} in terms of  standard deviation of the gravity coefficients estimate is clearly visible in Fig.~\ref{fig:spectre}. Despite the fact that such an improvement is expected following the addition of landmark data in our process, we think that these formal errors are overoptimistic because the large number $n$ of data points included in our fit (one million landmarks versus half a million Doppler measurements, see Tab.~\ref{tab:sum_data}) are assumed decorrelated (as commonly done) in the way they are computed (with a decrease following a $n^{-1/2}$ power law), which we know is not entirely true. Furthermore, we observe in this figure an inflection of the formal errors of degree 3 and 4 which is due to the fact that only a subset of coefficients at these degrees is estimated (see Tab.~\ref{tab:grav_field_deg_sup}).

\subsection{Motion of the center of mass}

Thanks to the addition of the landmarks, an accurate estimation of degree 1 coefficients is now possible and the shift between the CoR and the CoM can be inferred according to:
\begin{align*}
\Delta x &= \bar{C}_{11} R \sqrt{3} \\
\Delta y &= \bar{S}_{11} R \sqrt{3} \\
\Delta z &= \bar{C}_{10} R \sqrt{3}.    
\end{align*}
For each solution of degree 1 reported in Tab.~\ref{tab:grav_field_01}, we compute with the above equations the components of the CoR-CoM vector. Results are presented in Tab.~\ref{tab:com_displacement}. 
\begin{table*}[ht]
    \centering
    \caption{Amplitude in meters of the estimated shift between the centre of mass and the center of the Cheops reference frame. All uncertainties are formal errors ($1\sigma$). The two `Case 2/4` provide period-dependent estimates, while the others are global parameter estimates.}
    \label{tab:com_displacement}
    \begin{tabular}{l|c|c|c}
    \toprule
         & $\Delta x$ & $\Delta y$ & $\Delta z$  \\
        \midrule
        Shape (UMD)     &  $14.0$ & $-26.7$ &   $19.7$  \\
        Pätzold (D)     &  $0\pm 11$  & $4\pm 28$    & $-28\pm 37$ \\
        Godard (D\&L)   &       &    & $6.0\pm 0.7$ \\
        Case 0/2 (D\&L)         & $ 0.01\pm 0.002$ &  $0.011\pm 0.001$ & $ -7.6  \pm 0.1$ \\
        Case 2/4 Pre-P. (D\&L)  & $-0.054\pm 0.002$ &  $0.012\pm 0.002$ & $-35.443\pm 0.002$ \\
        Case 2/4 Post-P. (D\&L) & $ 0.075\pm 0.0008$ &  $0.034\pm 0.002$ & $  -0.207\pm 0.002$ \\                
        \bottomrule
    \end{tabular}
\end{table*}
Derived from different computations, one must be caution when comparing them, as they do not represent the same information. 
For instance, the `Shape (UMD)' shift corresponds to the vector between the CoR in which the body shape is described and the centre of gravity that the body would have if the distribution of masses were uniform, i.e. the CoR-CoF offset. The shift 'Pätzold (D)', inferred from RSI measurements only \citep{Patzold_2016}, corresponds to the CoR-CoM offset. Since RSI Doppler data are not physically linked to the body (as they only provide information about the projection of the spacecraft's velocity on the line of sight) the uncertainty on the CoR-CoM vector is very large.\\
Whereas the 'Case 0/2' only estimates a single value for degrees 1, which is a sort of average over all the fitted arcs, the 'Case 4/2' distinguishes between values before and after perihelion.  This allows us to observe a large variation in the comet's CoM location during its trajectory around the Sun. In particular,  the CoM moved along the Z axis by $+35.2\,m$. 
The loss of mass and the redeposition of dust may explain this observation. 
Indeed, the orientation of the comet as it passed through perihelion was such that the southern solstice took place 34 days after perihelion, implying a strong sunlight imbalance between the comet's northern and southern hemispheres. This should have resulted in a much more pronounced out-gassing in the south 
shifting the CoM northward.

\subsection{Correlations}

\begin{figure*}[!ht]
      \centering
      \includegraphics[width=0.68\textwidth, trim=0 30pt 0 10pt,clip]{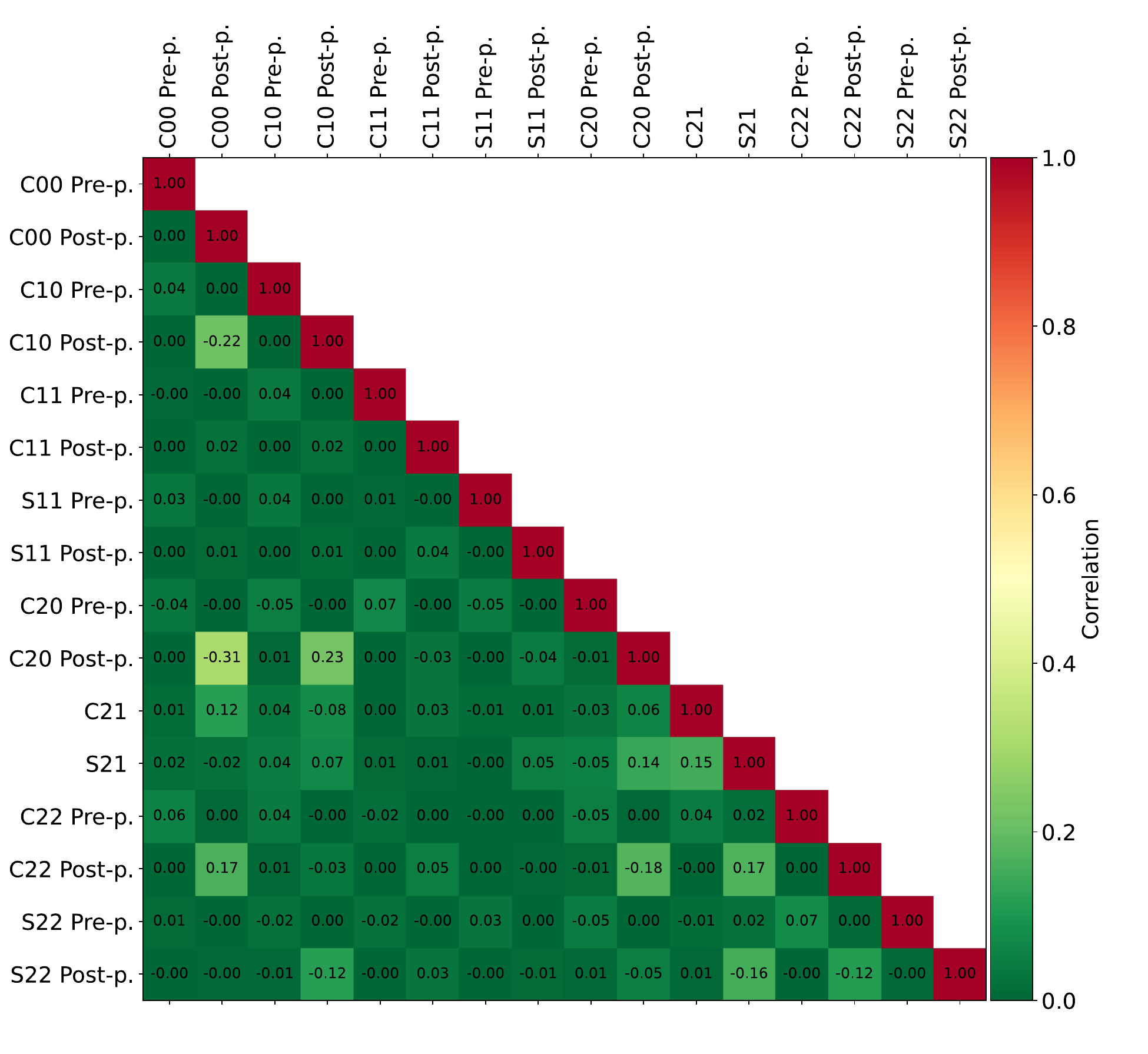}
      \includegraphics[width=0.68\textwidth, trim=0 30pt 0 10pt,clip]{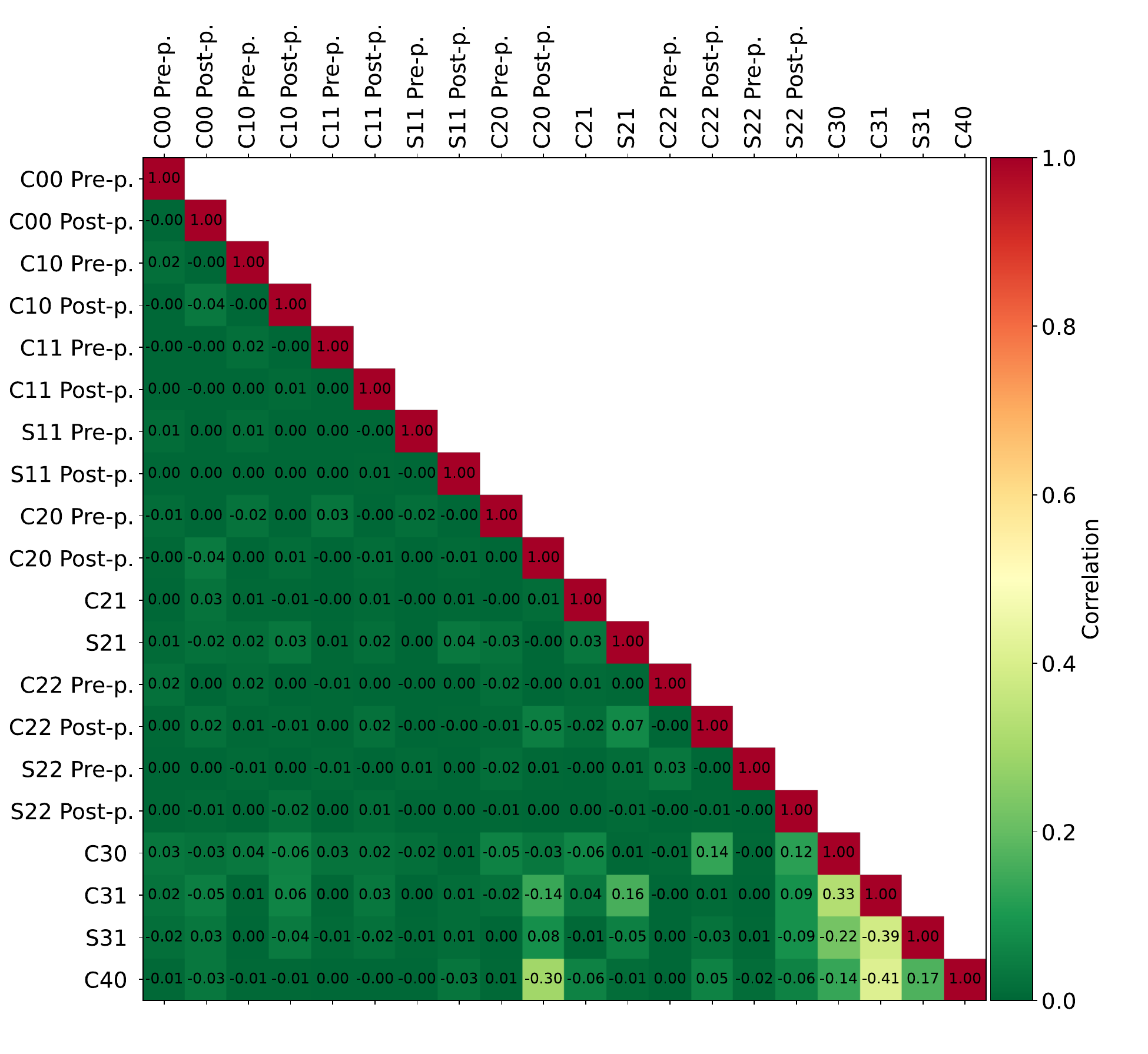}
      \caption{Correlation matrix for the gravity field coefficients on ‘step 1’ (top) and ‘step 2’ (bottom)} 
      \label{fig:correl}
 \end{figure*}

The correlations between the estimated gravity coefficients are shown in Fig.~\ref{fig:correl} for the ‘step 1’ and for the ‘step 2’ of our inversion procedure. A large majority of them are less than 0.1, which is quite satisfactory. For ‘step 1’, the largest correlations are between $\bar{C}_{00}$, $\bar{C}_{10}$ and $\bar{C}_{20}$ post-perihelion coefficients. However, they are in absolute value between 0.22 and 0.31, which is still quite small and acceptable. For ‘step 2’, as expected, the coefficients with the highest degrees are the most correlated (up to 0.41). In fact, the greater the number of parameters for the same number of measurements, the more correlated the solutions obtained will be, especially if the signature of the parameters added to the measurements decreases, which is the case with degree 3 and 4 coefficients, where we believe we have reached the limit of accessible precision. It should be noted that the correlations of the lowest degree coefficients are further reduced with respect to ‘step 1’, as a consequence of the resolution strategy that adds constraints on degrees lower than 2 at ‘step 2’.  

\subsection{Residuals}

RMS postfit residuals are good indicators of the quality of the fit and of the validity of the solutions. Those of Case 2/4 are provided in Tab.~\ref{tab:rms} per data type. They are satisfactory, with mean values just above the measurement noise. 

\begin{figure*}[ht]
     \centering
     \includegraphics[width=0.49\textwidth]{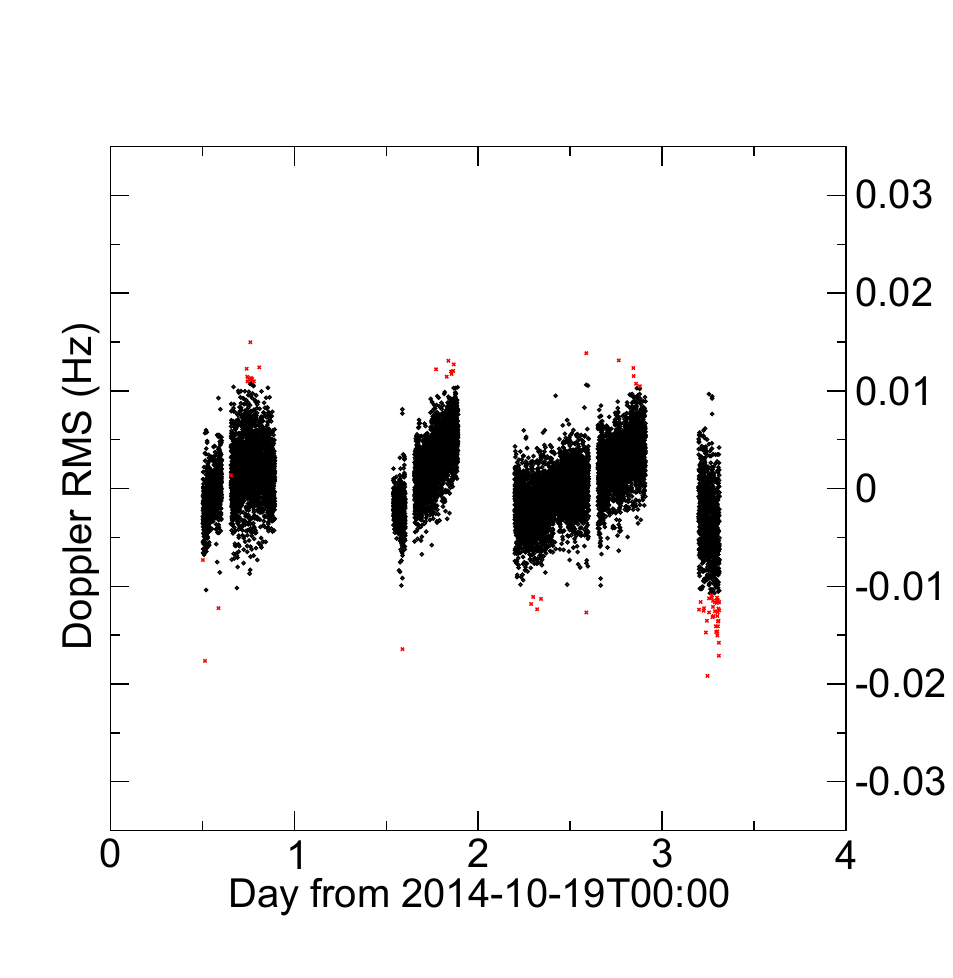}
     \includegraphics[width=0.49\textwidth]{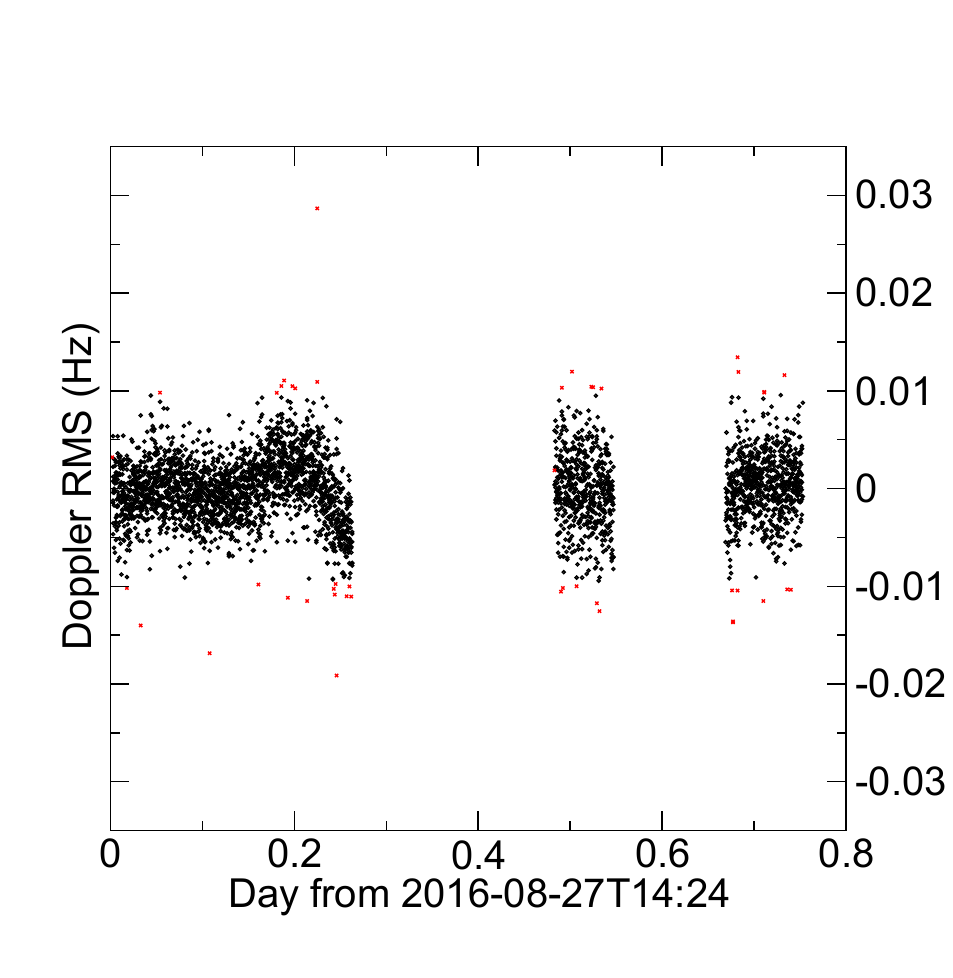}
     \caption{Doppler residuals of arc~\#8 pre-perihelion (left) and arc~\#23 post-perihelion (right) as a function of time. Red cross are rejected measurements, Black dot are kept measurements.}\label{fig:dopResArc}
\end{figure*}

\begin{figure*}[ht]
     \centering
     \includegraphics[width=0.49\textwidth]{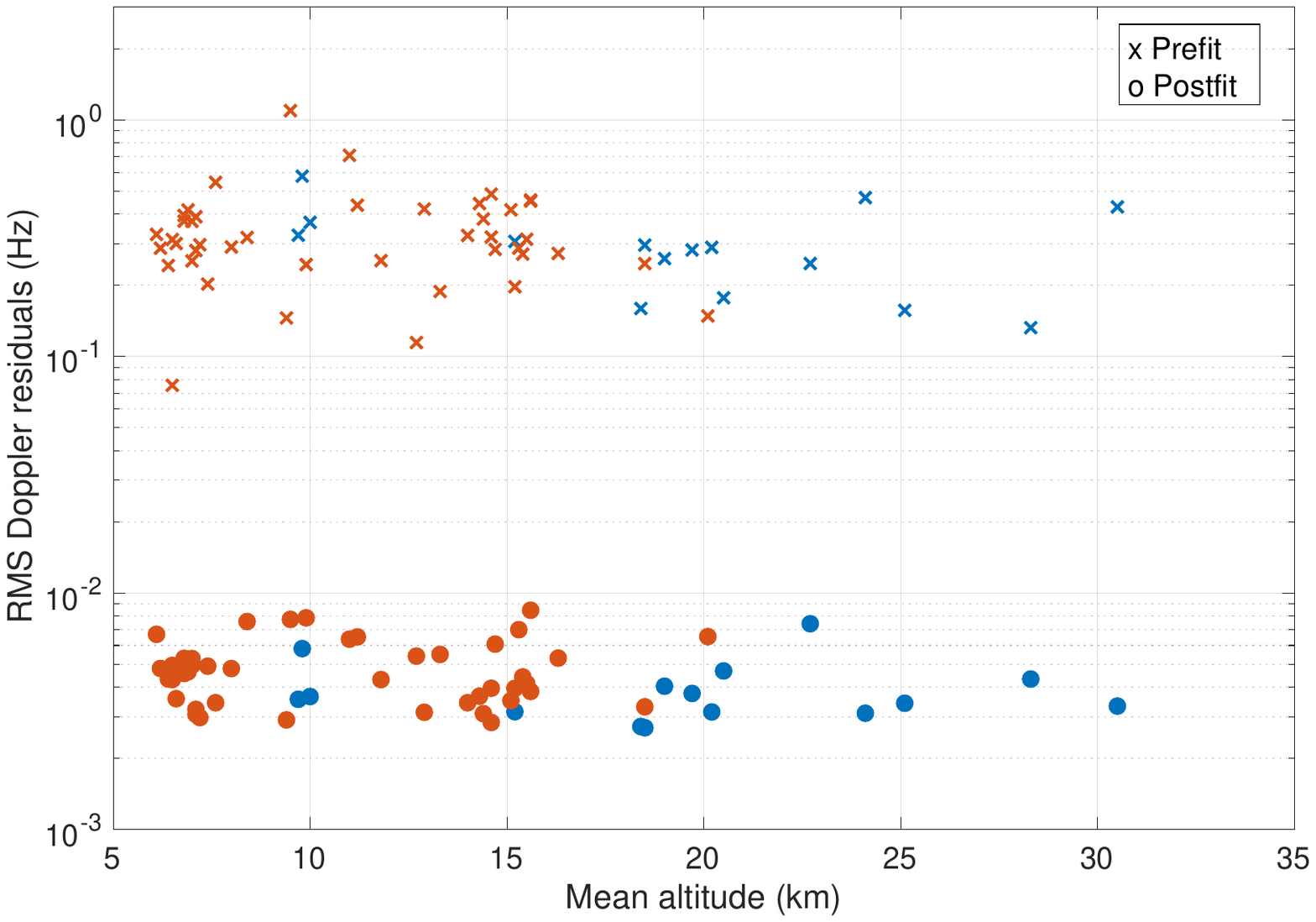}
     \includegraphics[width=0.49\textwidth]{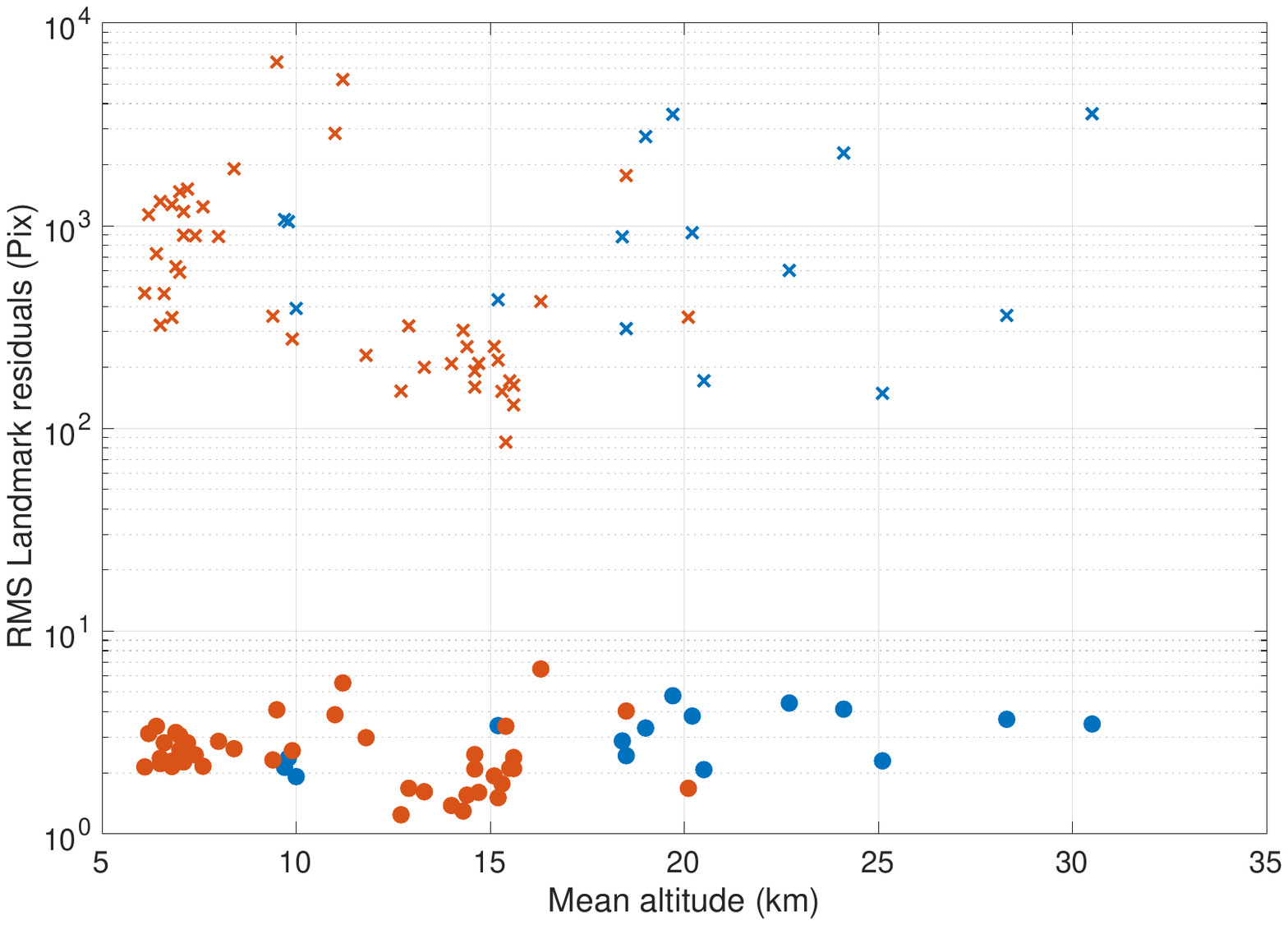}
     \caption{Prefit and Postfit RMS residuals of Doppler (left) and landmark (right) data collected before perihelion (blue) and after perihelion (orange) as a function of arcs mean altitude.}  \label{fig:dopResPrePost}
\end{figure*}

\begin{table}[ht]
    \centering
    \caption{RMS of the postfit residuals by measurement type obtained at convergence of Case 0/2 and Case 2/4. The minimum, mean and maximum RMS of the residuals per measurement type are distinguished. }
    \label{tab:rms}
    \begin{tabular}{c|c|c|c|c}
    
    \toprule
         Case & Data type & Min & Mean & Max  \\
    \midrule
         \multirow{ 2}{*}{Case 0/2} & Doppler (mHz)  & 2.7 & 5.1 & 21.6  \\
                                          & Landmarks (px) & 1.2 & 3.2 & 6.5  \\
    \midrule
         \multirow{ 2}{*}{Case 2/4} & Doppler (mHz)  & 2.8 & 4.5 & 8.5  \\
                                          & Landmarks (px) & 1.2 & 3.1 & 6.5  \\
    \bottomrule
    \end{tabular}
\end{table}
Fig.~\ref{fig:dopResArc} shows the residual profile of the Doppler measurements of arc~\#8 (representative of a long arc) before perihelion and arc~\#23 (representative of a short arc) after perihelion. The profiles are not perfectly flat (which is classical in POD), suggesting that there are still information in the data that our estimated solution doesn't fully represent. The tendencies observed can be explained by the comet's ephemeris, which is not adjusted in our case. An optimisation of the orbit of 67P/C-G could erase this last tendency. Nevertheless, Fig.~\ref{fig:dopResPrePost} clearly shows the decrease of the postfit residuals with respect to the prefit residuals, revealing the improvement of our adjusted dynamical model and trajectory with respect to our a priori model, based on the UMD gravity field. 

\section{Discussion}
\label{sec:discussion}

\subsection{Added value of optical measurements}

\begin{figure}[ht]
     \centering
     \includegraphics[width=\userfigwidth]{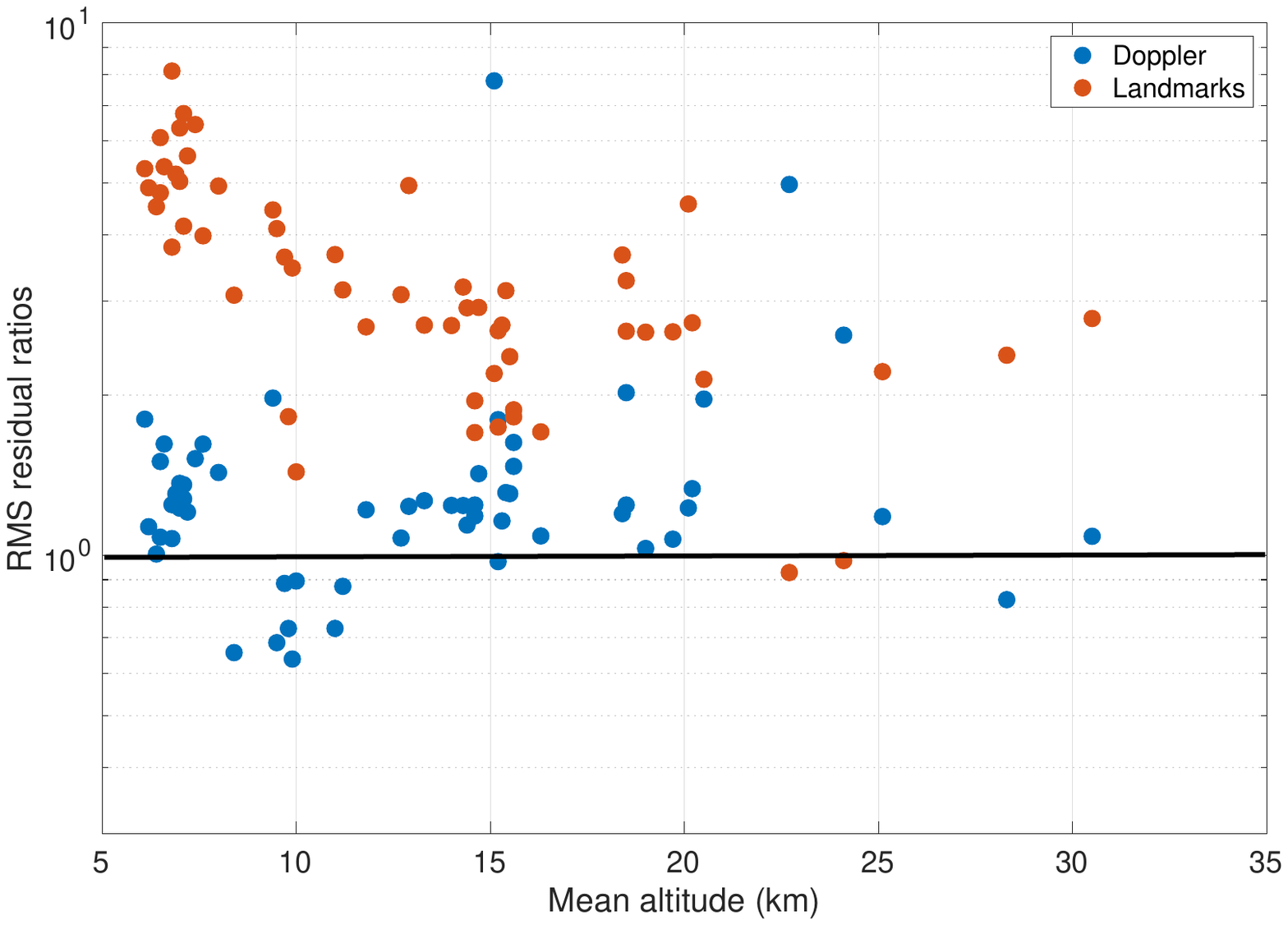}
     \caption{Data residual ratios between those obtained with the ESOC orbit and those obtained at convergence in this study (i.e. ESOC residuals / convergence residuals).}
     \label{fig:resRatios}
\end{figure}

We here discuss and quantify the added value of the landmark measurements since they represent the main difference with respect to previous studies. The first indicator of the beneficial addition of landmarks is the number of converging arcs that is approximately twice that without landmark data, resulting in better resolution and precision of the gravity solution. Secondly, one can see in the data residuals that the quality of our orbit and dynamical model as a whole has clearly benefited from the addition of landmarks. Indeed, besides the improvement of our estimated model and trajectory over our a priori ones (Fig.~\ref{fig:dopResPrePost}), the improvement of our orbit with respect to that of ESOC is more subtle, but nevertheless perceptible as shown in Fig.~\ref{fig:resRatios}. On this figure we have plotted the ratio between the residuals obtained with the orbit of ESOC and with ours. Above one, the ratio means our orbit matches the measurements better. As one can see, our landmark residuals are 2 to 10 times smaller than those calculated with the ESOC orbit. Although both solutions are obtained from a combination of Doppler and image data, this result was expected since ESOC did not use the OSIRIS landmarks as such. What is even more interesting is that the Doppler residual ratios, generally closer to one, also show a notable improvement in a large majority of low altitude orbits (the most sensitive to the gravity field), revealing the interest of including landmark data in the inversion process.   \\

\begin{figure*}[ht]
    \centering
    \includegraphics[width=\textwidth,trim=0 50pt 0 20pt,clip]{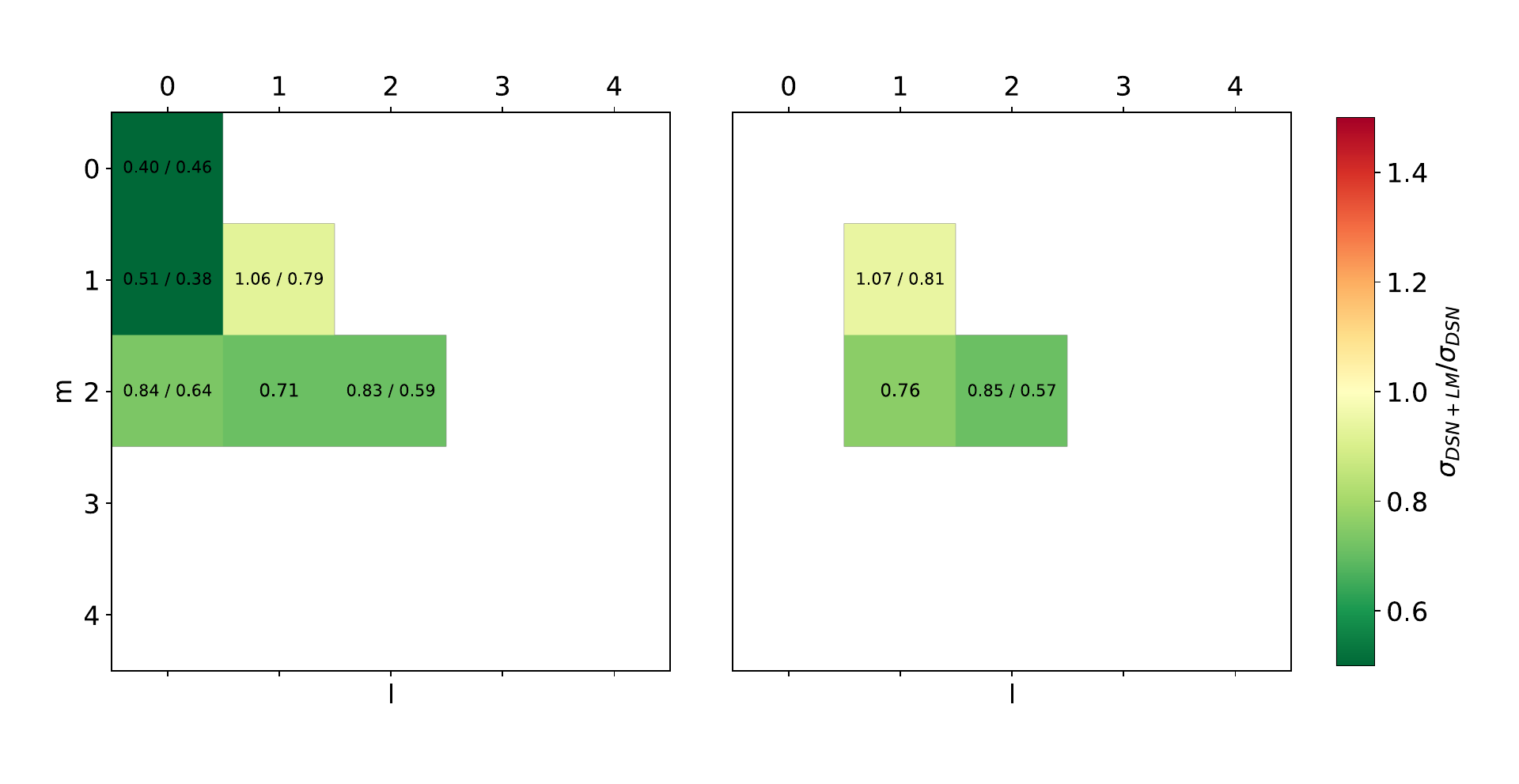}
    \includegraphics[width=\textwidth,trim=0 50pt 0 20pt,clip]{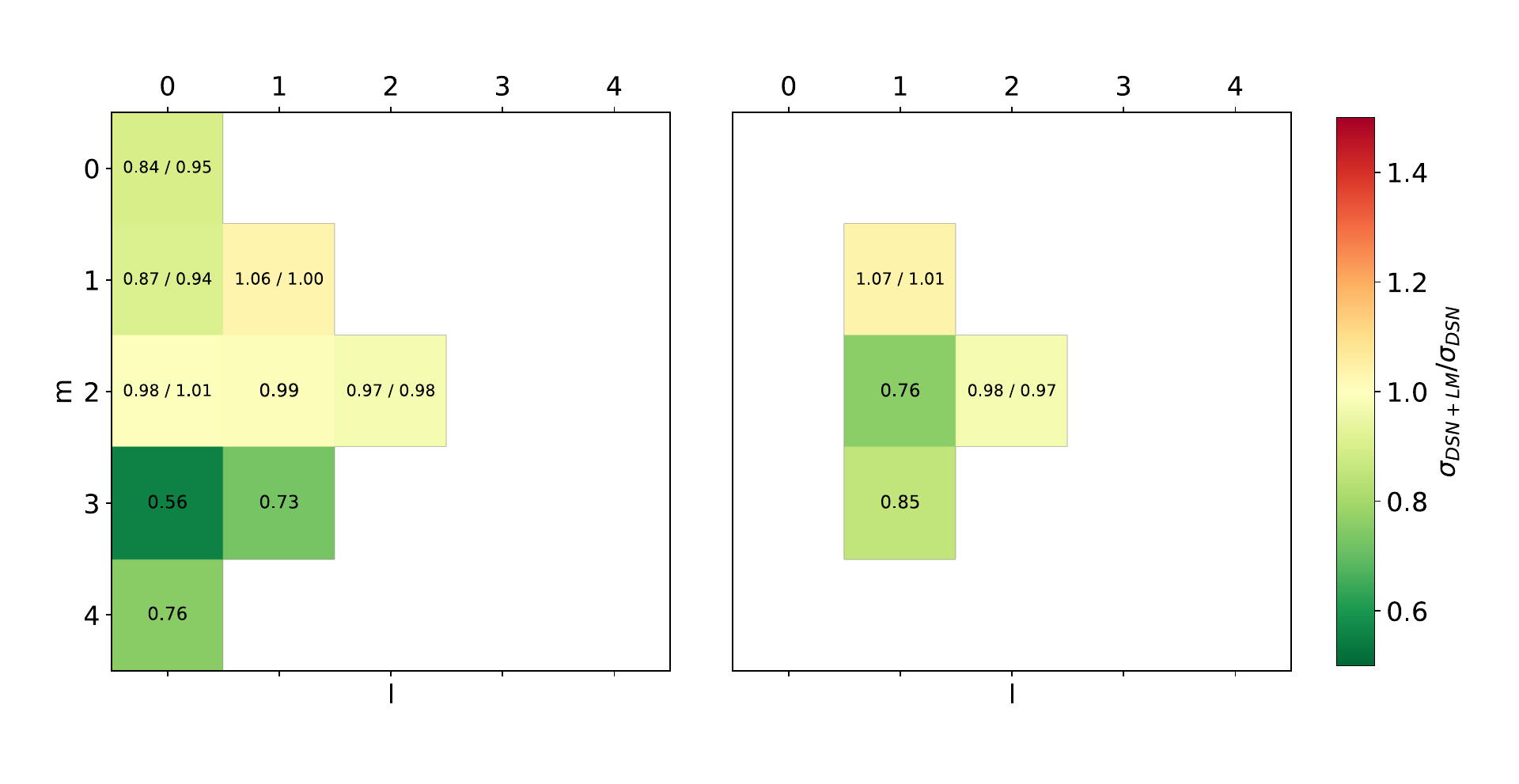}
    \caption{Standard deviation reduction factor from Doppler only to Doppler plus landmark data used in the estimation for $\bar{C}_{lm}$ (left panels) and $\bar{S}_{lm}$ (right panels) coefficients. Values smaller than one indicate an improvement resulting from the addition of landmarks. The two upper graphs show the results of ‘step 1’ and the two lower graphs show the results of ‘step 2’.} 
    \label{fig:sig_evol_C}
\end{figure*}

\noindent
At the point of convergence, it is possible to calculate the normal equations according to each contribution: one normal equation for the Doppler measurements only, and one normal equation for the Doppler plus landmark measurements. An inversion of these two equations allows us to assess the linear impact of the landmark addition on the solution and its accuracy. Both calculations leads to parameter values that are statistically equivalent, but the standard deviations and correlations are very different, mostly because the number of measurements is drastically different between the two cases ($\sim 470k$ Doppler measurements for both and $\sim 1M$ extra landmark measurements for the combined case), but also because the two types of data are complementary, so with different sensitivities to the estimated parameters. This is illustrated on Fig.~\ref{fig:sig_evol_C}, which shows the ratios per Stokes coefficient between the standard deviations of the Doppler plus landmark case and those of the Doppler only case, for ‘step 1’ and ‘step 2’ separately.



For ‘step~1’, a strong impact of landmark measurements on the standard deviation of $\bar{C}_{00}$ and $\bar{C}_{10}$ is observed, whereas $\bar{C}_{11}$ and $\bar{S}_{11}$ are less dependent on the addition of landmark observations. This is because the latter are already well constrained by the dynamics induced by degree~1 (related to the non-inertial frame acceleration, see Sec.~\ref{sec:deg1}). The uncertainties of the degree~2 coefficients are also improved by the landmark measurements, but to a lesser extent than those of the $\bar{C}_{00}$ and $\bar{C}_{10}$. 

For ‘step~2’, the standard deviation of coefficients of degree $\le2$ are constrained by the resolution strategy, so the impact of landmark measurements is only visible on the fully freed coefficients of degree~3 and 4. We observe a very strong improvement on the uncertainty of $\bar{C}_{30}$, which is linked to the north/south asymmetry and is clearly visible thanks to the landmarks. $\bar{C}_{40}$, which represents information that is averaged over one orbit revolution, and is therefore slightly less observable using a POD enhanced by the landmarks.


\subsection{Bulk density and porosity}

\citep{Patzold_2019} interpreted their GM estimate in terms of bulk density, deducing limits on the porosity of the material making up the comet. As our estimate of GM is close to theirs (666.1 $m^3/s^2$ compared with 666.2 $m^3/s^2$), the estimate of the mean density and resulting constraints on the porosity remain unchanged.

\subsection{Implications of mass loss}

\begin{figure*}[ht]
     \centering
     \includegraphics[width=\userfigwidth]{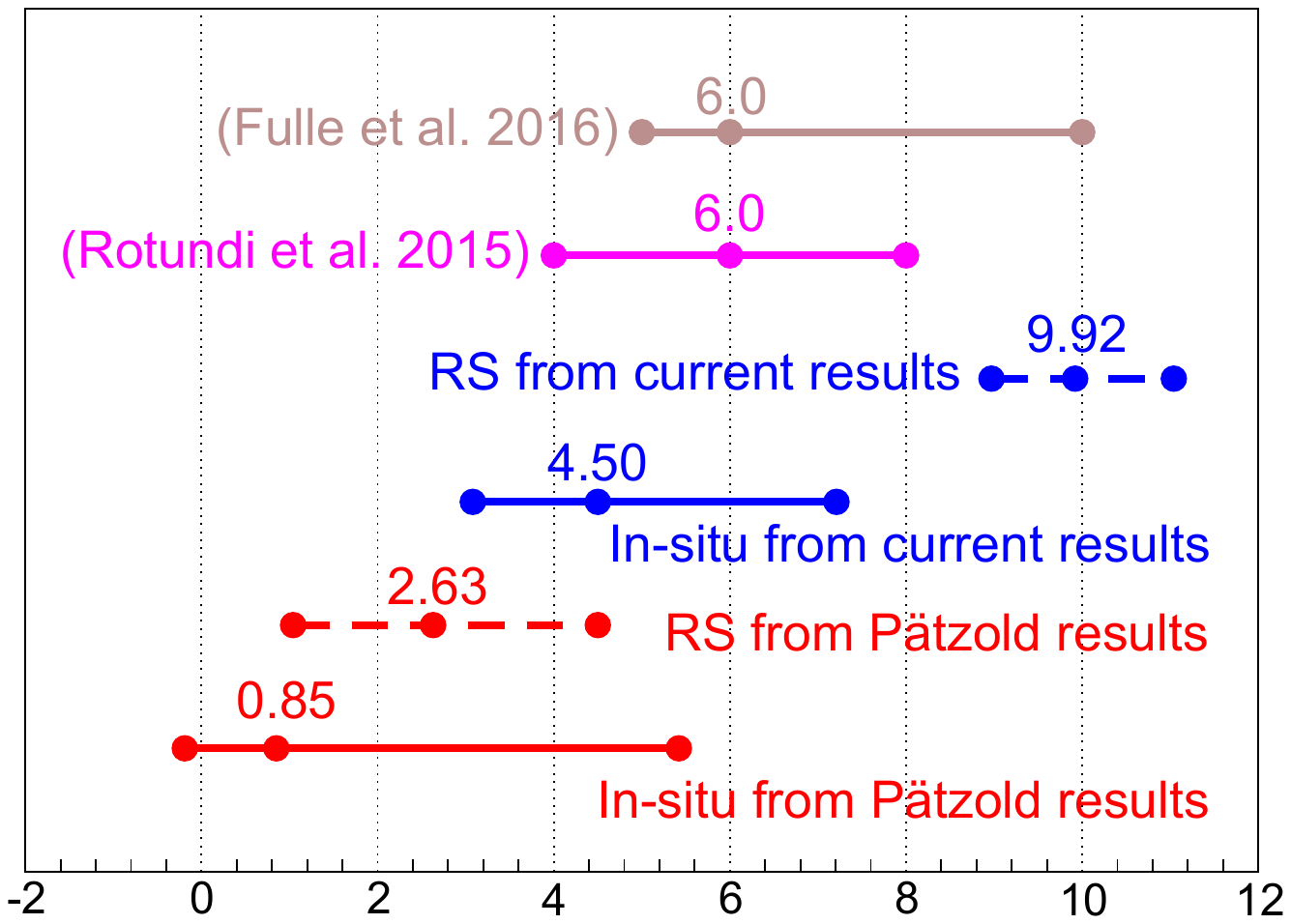}
     \includegraphics[width=\userfigwidth]{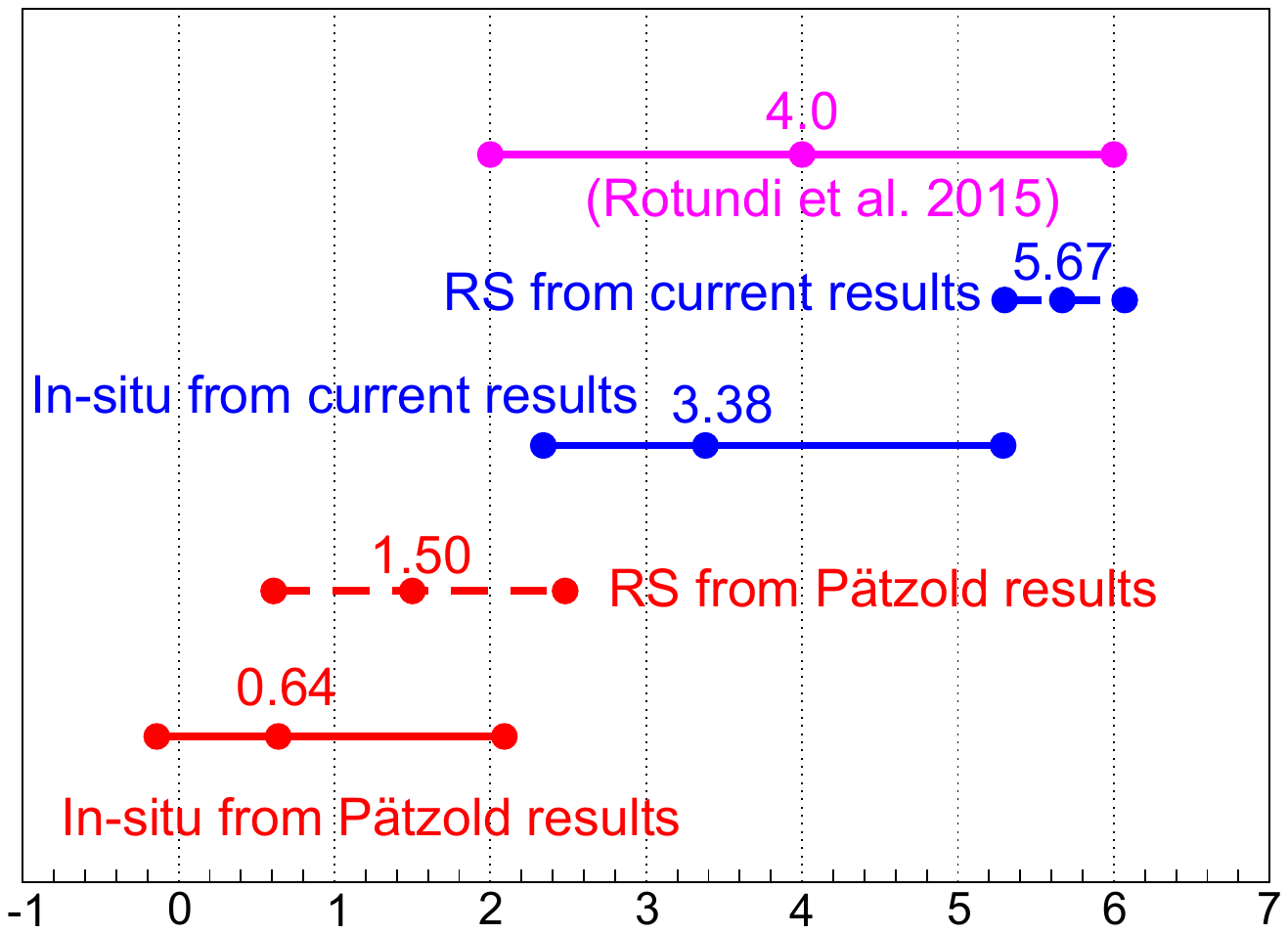}
     \caption{Ranges of plausible values derived for the dust-to-water (left) dust-to-all-volatiles (right) ratios. RS stands for remote sensing.}
     \label{fig:delta_DG_H20_V}
\end{figure*}

One of the main results of this study is the new mass loss ($\Delta M$) estimate that is rather different from the value published by \cite{Patzold_2019}, most likely due to our introduction of the frame acceleration ($\vec{\gamma_1}$). 
The $\Delta M$ of \cite{Patzold_2019} was used by \cite{Choukroun_2020} to determine the dust-to-gas ratio ($\delta_{DG}^V$) in the coma. However, the value they obtained  (i.e. $\delta_{DG}^V<1$  for all volatiles) is incompatible with measurements from GIADA (see \cite{Rotundi2015}), which reported $\delta_{DG}^V = 4 \pm 2$. We have reproduced the calculations carried out in \cite{Choukroun_2020} and made similar figures representing the ranges of plausible values obtained for the dust-to-gas ratio. Fig.~\ref{fig:delta_DG_H20_V} shows the values that we obtain based on our comet mass loss estimate for the dust-to-water and dust-to-all-volatiles ratios, superimposed to those previously published. The ratios are obtained by combining our $\Delta M$ value with the loss of volatile elements, which can be observed in two ways: either in-situ or through remote sensing (see \cite{Choukroun_2020} for more details).

These two types of observations result in fundamentally incompatible values for the dust-to-water and dust-to-all-volatiles ratios. Although this intrinsic discrepancy makes it hard to reach a firm conclusion, one can see in fig. \ref{fig:delta_DG_H20_V} that the values derived from our $\Delta M$ estimate are overall closer to GIADA measurements than those derived from the $\Delta M$ previously given in \cite{Patzold_2019}. The ratio based on in-situ data are in good agreement with the previous solutions of \cite{Rotundi2015} and \cite{Fulle2016}, and, to a lesser extent with those deduced by \cite{Choukroun_2020} using the $\Delta M $ from \cite{Patzold_2019}. These new findings could help put a better constraint on the dust-to-gas ratios.

\subsection{(Non)-homogeneity of the nucleus}
As demonstrated in Sec.~\ref{sec:results}, our estimated gravity field is different from that computed under uniform density assumption. In particular, a clear offset (several orders of magnitude larger than $1\sigma$) is observed between the center of reference and the center of mass, revealing a now incontestable level of heterogeneity in the comet. The origin of such heterogeneity can be multiple, but the fact that we observed a displacement of the CoM away from the direction of the Sun during the passage of the comet at perihelion, leads us to believe that ice sublimation is probably the mechanism responsible for this heterogeneity.\\

\noindent
Just like degree-1 coefficients, the degree-2 coefficients can also be used to constrain the interior of the comet as they depend on its moments of inertia according to the following expressions

\begin{align*}
\bar{C}_{20} M R^2 \sqrt{5} &= \frac{I_{xx}+I_{yy}}{2} - I_{zz} \\
\bar{C}_{21} M R^2 \sqrt{\frac{5}{3}} &= I_{xz} \\
\bar{S}_{21} M R^2 \sqrt{\frac{5}{3}} &= I_{yz} \\
\bar{C}_{22} M R^2 \sqrt{\frac{5}{12}} &= \frac{I_{yy}-I_{xx}}{4} \\
\bar{S}_{22} M R^2 \sqrt{\frac{5}{12}} &= \frac{I_{xy}}{2}.
\end{align*}
The retrieval of the moments of inertia from degree-2 is thus possible in theory, although this is an ill-posed problem because one cannot univocally determine the six moments of inertia ($I_{xx},I_{yy},I_{zz},I_{xy},I_{xz},I_{yz}$) from the five estimated degree-2 coefficients. \\
When no additional information is available, one can interpret the estimated coefficients in terms of interior properties simply by comparing them to the coefficients computed under UMD assumption. For the degree-2 specifically, we cross validate our own computation based on the Shape model with that inferred from the above equations using the complete inertia matrix provided by \cite{Kramer2019} (see their Eq.~(18)) under similar UMD assumption. The result of such calculations is reported in Tab.~\ref{tab:grav_kramer}. 
The very good agreement between these two sets of coefficients provides a robust reference against which estimated degree-2 parameters can be compared to discuss the level of heterogeneity in the comet. \\
In the case of the 67P, we are fortunate to have additional information regarding its internal mass distribution. Indeed, as mentioned in the introduction, \cite{Gutierrez_2016} concluded from the analysis of the rotational motion of 67P that the interior of the comet should not be homogeneous. An interesting outcome of their study are the two linear equations (Eq.10 in their paper) relating the main moments of inertia to each other. Combined with the above equations relating $\bar{C}_{20}$ and $\bar{C}_{22}$ to $I_{xx},I_{yy}$ and $I_{zz}$, one can univocally return to $I_{xx},I_{yy},I_{zz}$ from our estimates of $\bar{C}_{20}$ and $\bar{C}_{22}$. Accounting from the uncertainties in the model parameters of \cite{Gutierrez_2016} (i.e. the difference between the coefficients of their two linear equations) and in our gravity estimates (at 10$\sigma$), we get the following ranges of moments of inertia for 67P:
\begin{align*}
    I_{xx} / (M R^2) & \in [0.00344,0.03973],\\
    I_{yy} / (M R^2) & \in [0.12170,0.15798],\\
    I_{zz} / (M R^2) & \in [0.12355,0.19141].       
\end{align*}
 As expected given the difference between our degree-2 estimated coefficients and the one reported in Tab.~\ref{tab:grav_kramer}, these 10$\sigma$ ranges of moment of inertia do not include their homogeneous counterparts ($I^{h}_{xx} / (M R^2) = 0.13607;I^{h}_{yy} / (M R^2) = 0.25115, I^{h}_{zz} / (M R^2) = 0.27031$), proving from another point of view (parameter-wise) the existence of large-scale heterogeneities inside the comet. A companion paper based on the method of \cite{Caldiero2024} will discuss in more detail the inference of the internal properties of 67P from our new gravity field and the rotation state of the comet.

\begin{table}[ht]
\caption{Comparison of degree-2 coefficients calculated from the shape under the assumption of uniform mass distribution, along with the reinterpretation of the inertia matrices from Kramer's publication. Both are consistent with the CHEOPS reference frame.}\label{tab:grav_kramer}
\begin{tabular}{cccccc}
\toprule
  & $\bar{C}_{20}$ & $\bar{C}_{21}$ & $\bar{S}_{21}$ & $\bar{C}_{22}$ & $\bar{S}_{22}$ \\
\midrule
        Shape  & -0.0342 & -0.0026 & 0.0008 & 0.0445 & -0.0001 \\
        Kramer & -0.0343 & -0.0025 & 0.0008 & 0.0446 & -0.0004 \\
\bottomrule
\end{tabular}
\end{table}

\section{Conclusion}
\label{sec:conclusion}
In this study we have reevaluated the gravity field of comet 67P/C-G by using optical observations in addition to traditional Doppler measurements. Thanks to the complementary of these types of data, both the accuracy and resolution of the gravity field of the comet are significantly improved. The new field resulting from this analysis is estimated with statistical significance up to degree 4. Consistent with the previous solutions, the order-of-magnitude more precise field we obtain here allows us to detect heterogeneities in the comet's nucleus that were not observed before from the less precise field. Two major results emerge from our analysis. The first is that we estimate a mass loss due to ice sublimation at perihelion that is 2.8 times larger than previously estimated \citep{Patzold_2019}. This leads to dust-to-water and dust-to-gaz ratios that are in better agreement with those measured with GIADA than before, which may have significant implications, especially for the composition of the coma. The second major result is that we observe, for the first time, a displacement of the center of mass of the comet during its flyby of the Sun. Inferred from a precise determination of the degree-1 gravity coefficients and their variations between pre and post perihelion, this northward shift of $\sim35\,m$ could be explained by a more pronounced outgassing activity in the south of the comet than in the north, due to the orientation of its spin axis relative to the Sun.\\
Finally, this study highlights the benefits of combining radiometric and landmark-based techniques to better estimate geodetic parameters of small bodies. More generally, the use of positional anchors as landmarks and/or altimetry data (e.g. LIDAR) could prove essential for the precise orbit determination of spacecraft around small bodies with the ultimate goal of probing their interior (e.g. \cite{Gramigna:2024ab}).

\section{CRediT authorship contribution statement}
{ \bf Julien Laurent-Varin}: Conceptualization, Methodology, Software, Formal analysis, Writing - Review \& Editing, Visualization, Supervision
{ \bf Théo James}: Investigation, Methodology, Software, Writing - Original Draft, Writing - Review \& Editing, Visualization
{ \bf Jean-Charles Marty}: Software, Methodology, Writing - Review \& Editing
{ \bf Laurent Jorda}: Resources, Writing - Review \& Editing
{ \bf Sebastien Le Maistre}: Resources, Writing - Review \& Editing, Visualization, Formal analysis
{ \bf Robert Gaskell}: Resources, Writing - Review \& Editing

\section{Declaration of competing interest}
The authors declare that there is no competing interest...

\section{Acknowledgments}
The authors thanks A. Caldiero for his help and constructive discussions regarding the shape gravity field.

\newpage
\onecolumn

\appendix
\section{Arcs definition} \label{sec:arcs}

The Tabs.~\ref{tab:arcs_def_14} and \ref{tab:arcs_def_16} show the details of each arc chosen for the calculations. For each arc, the start and end dates and the duration are given. The root mean square residuals are also provided for the Doppler and landmark measurements for the last resolution of the 'Case 2/4'. Finally, the minimum, mean and maximum distances are provided for each arc.

\begin{table}[h!]
\caption{Arcs details before perihelion} \label{tab:arcs_def_14}
\begin{tabular*}{\tblwidth}{@{}C|CC|C|CC|CCC@{}}
\toprule
   & \multicolumn{2}{C|}{Date} &       & \multicolumn{2}{C|}{R.M.S.} & \multicolumn{3}{C}{Dist. to 67P/C-G center} \\
 N & Start      & End           & Len.  & Dop.     & Lnd.              & Min  & Mean & Max \\
   &            &               & [day] & [mHz]    & [Px]              & [km] & [km] & [km] \\
\midrule
 01 & 2014-09-21T12:02:36 & 2014-09-23T23:41:27 &  2.49 &        4.324 &        3.669 &      27.7 &      28.3 &      28.5 \\ 
 02 & 2014-09-24T10:22:35 & 2014-09-29T07:17:22 &  4.87 &        3.100 &        4.116 &      19.0 &      24.1 &      28.2 \\ 
 03 & 2014-09-29T10:22:35 & 2014-10-01T09:52:39 &  1.98 &        2.724 &        2.869 &      18.1 &      18.4 &      19.2 \\ 
 04 & 2014-10-01T13:18:36 & 2014-10-05T09:09:05 &  3.83 &        4.032 &        3.322 &      18.5 &      19.0 &      19.2 \\ 
 05 & 2014-10-05T09:13:06 & 2014-10-08T07:55:32 &  2.95 &        2.691 &        2.427 &      18.4 &      18.5 &      18.9 \\ 
 06 & 2014-10-12T09:13:06 & 2014-10-15T09:09:05 &  3.00 &        3.146 &        3.415 &       9.8 &      15.2 &      18.7 \\ 
 07 & 2014-10-15T09:13:06 & 2014-10-19T07:55:00 &  3.95 &        3.647 &        1.913 &       9.9 &      10.0 &      10.2 \\ 
 08 & 2014-10-19T12:02:36 & 2014-10-22T07:54:52 &  2.83 &        3.551 &        2.126 &       9.2 &       9.7 &      10.2 \\ 
 09 & 2014-10-22T14:24:00 & 2014-10-28T00:00:00 &  5.40 &        5.817 &        2.364 &       9.7 &       9.8 &       9.9 \\ 
 10 & 2014-11-04T00:00:00 & 2014-11-08T19:54:52 &  4.83 &        3.323 &        3.479 &      29.7 &      30.5 &      32.3 \\ 
 11 & 2014-12-03T05:54:18 & 2014-12-05T12:02:43 &  2.26 &        3.416 &        2.288 &      20.6 &      25.1 &      30.0 \\ 
 12 & 2014-12-06T02:22:38 & 2014-12-09T19:11:47 &  3.70 &        3.138 &        3.807 &      20.1 &      20.2 &      20.5 \\ 
 13 & 2014-12-10T04:44:13 & 2014-12-14T12:00:00 &  4.30 &        3.759 &        4.794 &      19.2 &      19.7 &      20.6 \\ 
 14 & 2014-12-17T04:02:38 & 2014-12-19T17:51:03 &  2.58 &        4.679 &        2.071 &      20.4 &      20.5 &      20.6 \\ 
 15 & 2014-12-20T04:02:38 & 2014-12-24T01:09:05 &  3.88 &        7.410 &        4.414 &      19.9 &      22.7 &      25.5 \\ 
\bottomrule
\end{tabular*}
\end{table}

\begin{table}[h!]
\caption{Arcs details after perihelion} \label{tab:arcs_def_16}
\begin{tabular*}{\tblwidth}{@{}C|CC|C|CC|CCC@{}}
\toprule
   & \multicolumn{2}{C|}{Date} &       & \multicolumn{2}{C|}{R.M.S.} & \multicolumn{3}{C}{Dist. to 67P/C-G center} \\
 N & Start      & End           & Len.  & Dop.     & Lnd.              & Min  & Mean & Max \\
   &            &               & [day] & [mHz]    & [Px]              & [km] & [km] & [km] \\
\midrule
 01 & 2016-04-27T01:46:39 & 2016-05-03T00:00:00 &  5.93 &        3.299 &        4.030 &      18.3 &      18.5 &      18.8 \\ 
 02 & 2016-05-08T06:31:09 & 2016-05-11T00:04:59 &  2.73 &        5.302 &        6.499 &      10.1 &      16.3 &      18.9 \\ 
 03 & 2016-05-11T06:31:09 & 2016-05-15T01:17:08 &  3.78 &        7.847 &        2.568 &       9.8 &       9.9 &      10.0 \\ 
 04 & 2016-05-15T07:30:10 & 2016-05-18T11:25:46 &  3.16 &        2.905 &        2.312 &       8.8 &       9.4 &       9.9 \\ 
 05 & 2016-05-18T16:22:58 & 2016-05-21T03:00:00 &  2.44 &        7.572 &        2.627 &       7.3 &       8.4 &       9.2 \\ 
 06 & 2016-05-22T12:00:00 & 2016-05-25T00:04:58 &  2.50 &        3.434 &        2.152 &       6.9 &       7.6 &       8.3 \\ 
 07 & 2016-05-25T06:31:09 & 2016-05-29T12:00:00 &  4.23 &        5.280 &        3.049 &       6.8 &       7.0 &       7.1 \\ 
 08 & 2016-07-20T05:12:37 & 2016-07-26T12:00:00 &  6.28 &        7.731 &        4.088 &       8.9 &       9.5 &      10.0 \\ 
 09 & 2016-07-27T03:38:01 & 2016-07-30T22:43:34 &  3.80 &        6.517 &        5.544 &       8.2 &      11.2 &      14.2 \\ 
 10 & 2016-07-31T03:25:22 & 2016-08-03T00:05:01 &  2.86 &        4.302 &        2.977 &       8.4 &      11.8 &      13.8 \\ 
 11 & 2016-08-03T06:31:10 & 2016-08-07T00:00:00 &  3.73 &        6.377 &        3.866 &       8.3 &      11.0 &      13.5 \\ 
 12 & 2016-08-10T11:18:13 & 2016-08-12T04:05:08 &  1.70 &        3.129 &        1.675 &      10.7 &      12.9 &      13.8 \\ 
 13 & 2016-08-12T11:18:13 & 2016-08-13T05:22:10 &  0.75 &        4.789 &        2.856 &       7.5 &       8.0 &       9.0 \\ 
 14 & 2016-08-13T12:58:37 & 2016-08-15T06:35:07 &  1.73 &        5.406 &        1.240 &      10.4 &      12.7 &      13.7 \\ 
 15 & 2016-08-15T11:18:11 & 2016-08-16T05:22:10 &  0.75 &        4.901 &        2.447 &       6.7 &       7.4 &       8.9 \\ 
 16 & 2016-08-16T12:58:37 & 2016-08-18T06:12:38 &  1.72 &        5.496 &        1.611 &      10.8 &      13.3 &      14.5 \\ 
 17 & 2016-08-18T11:18:11 & 2016-08-19T05:22:10 &  0.75 &        3.067 &        2.260 &       6.0 &       7.1 &       8.9 \\ 
 18 & 2016-08-19T12:58:37 & 2016-08-21T04:26:51 &  1.64 &        3.434 &        1.378 &      11.5 &      14.0 &      15.1 \\ 
 19 & 2016-08-21T11:18:11 & 2016-08-22T05:22:10 &  0.75 &        4.565 &        2.145 &       5.5 &       6.8 &       9.1 \\ 
 20 & 2016-08-22T12:58:37 & 2016-08-24T04:27:39 &  1.65 &        3.087 &        1.551 &      11.7 &      14.4 &      15.6 \\ 
 21 & 2016-08-24T11:18:11 & 2016-08-25T09:03:58 &  0.91 &        2.972 &        2.807 &       5.0 &       7.2 &      10.4 \\ 
 22 & 2016-08-25T09:08:00 & 2016-08-27T07:19:59 &  1.92 &        3.658 &        1.294 &      10.4 &      14.3 &      16.0 \\ 
 23 & 2016-08-27T11:18:11 & 2016-08-28T09:03:59 &  0.91 &        3.219 &        2.656 &       4.7 &       7.1 &      10.7 \\ 
 24 & 2016-08-28T09:08:01 & 2016-08-30T05:55:09 &  1.87 &        2.835 &        2.086 &      10.6 &      14.6 &      16.3 \\ 
 25 & 2016-08-30T11:18:13 & 2016-08-31T05:22:36 &  0.75 &        4.303 &        2.218 &       4.4 &       6.5 &       9.4 \\ 
 26 & 2016-08-31T12:58:37 & 2016-09-02T05:12:39 &  1.68 &        3.505 &        1.928 &      12.2 &      15.1 &      16.4 \\ 
 27 & 2016-09-02T12:18:11 & 2016-09-03T05:22:10 &  0.71 &        4.793 &        3.121 &       4.1 &       6.2 &       9.0 \\ 
 28 & 2016-09-03T12:58:37 & 2016-09-05T07:07:20 &  1.76 &        8.451 &        2.096 &      12.1 &      15.6 &      17.1 \\ 
 29 & 2016-09-05T11:18:11 & 2016-09-06T08:54:07 &  0.90 &        4.944 &        2.591 &       3.9 &       7.0 &      10.8 \\ 
 30 & 2016-09-06T12:58:38 & 2016-09-08T07:19:59 &  1.76 &        3.833 &        2.377 &      12.6 &      15.6 &      17.1 \\ 
 31 & 2016-09-08T11:18:12 & 2016-09-09T05:22:10 &  0.75 &        3.571 &        2.809 &       4.1 &       6.6 &      10.7 \\ 
 32 & 2016-09-09T12:58:38 & 2016-09-11T07:20:01 &  1.76 &        3.953 &        2.458 &      11.4 &      14.6 &      16.1 \\ 
 33 & 2016-09-11T11:18:11 & 2016-09-12T05:22:10 &  0.75 &        4.324 &        3.388 &       4.1 &       6.4 &       9.8 \\ 
 34 & 2016-09-12T11:05:38 & 2016-09-14T04:23:22 &  1.72 &        4.414 &        3.390 &      11.4 &      15.4 &      16.8 \\ 
 35 & 2016-09-14T11:18:11 & 2016-09-15T07:19:59 &  0.83 &        5.288 &        2.278 &       4.1 &       6.8 &      10.8 \\ 
 36 & 2016-09-15T12:58:38 & 2016-09-17T04:24:03 &  1.64 &        6.077 &        1.599 &      11.4 &      14.7 &      16.1 \\ 
 37 & 2016-09-17T11:18:11 & 2016-09-18T09:03:57 &  0.91 &        4.637 &        3.158 &       4.1 &       6.9 &      11.1 \\ 
 38 & 2016-09-18T09:07:57 & 2016-09-20T06:45:32 &  1.90 &        6.993 &        1.764 &      11.0 &      15.3 &      17.0 \\ 
 39 & 2016-09-20T11:18:11 & 2016-09-21T05:22:10 &  0.75 &        4.939 &        2.356 &       4.1 &       6.5 &      10.2 \\ 
 40 & 2016-09-21T12:58:38 & 2016-09-23T04:23:34 &  1.64 &        4.141 &        2.103 &      12.8 &      15.5 &      16.7 \\ 
 41 & 2016-09-23T11:42:33 & 2016-09-24T05:22:10 &  0.74 &        6.685 &        2.136 &       4.1 &       6.1 &       9.2 \\ 
 42 & 2016-09-24T12:55:26 & 2016-09-26T04:23:20 &  1.64 &        3.950 &        1.506 &      11.2 &      15.2 &      17.2 \\ 
 43 & 2016-09-26T12:49:05 & 2016-09-28T00:02:38 &  1.47 &        6.538 &        1.675 &      17.5 &      20.1 &      22.2 \\ 
\bottomrule
\end{tabular*}
\end{table}


\newpage
\twocolumn
\bibliography{biblio}


\end{document}